\documentclass[twocolumn]{aastex63}

\newcommand{\be}{\begin{equation}}
\newcommand{\ee}{\end{equation}}

\defcitealias{Mignani2010}{MPK10}

\usepackage{amssymb,amsmath}
\usepackage{color}

\shorttitle{XMM and HST observations of PSR\,B1055$-$52}
\shortauthors{Posselt et al.}
\begin{document}

\title{X-ray and near-infrared observations of the middle-aged pulsar B1055--52, its multiwavelength spectrum, and proper motion\footnote{Based on observations obtained with XMM-Newton, an ESA science mission with instruments and contributions directly funded by ESA Member States and NASA. Based also on observations made with the NASA/ESA {\sl Hubble Space Telescope}, obtained at the Space Telescope Science Institute, which is operated by the Association of Universities for Research in Astronomy, Inc., under NASA contract NAS 5-26555. These observations are associated with program \#15676.}}

\correspondingauthor{B. Posselt}
\email{bettina.posselt@physics.ox.ac.uk}

\author[0000-0003-2317-9747]{B. Posselt}
\affiliation{Oxford Astrophysics, University of Oxford,
Denys Wilkinson Building, Keble Road, 
Oxford OX1 3RH, UK
}
\affiliation{Department of Astronomy \& Astrophysics, Pennsylvania State University, 
525 Davey Lab, 
16802 University Park, PA, USA
}

\author[0000-0002-7481-5259]{G. G. Pavlov}
\affiliation{Department of Astronomy \& Astrophysics, Pennsylvania State University,
525 Davey Lab,
16802 University Park, PA, USA
}
\author[0000-0002-6447-4251]{O. Kargaltsev}
\affiliation{ Department of Physics, George Washington University, 
725 21st Street NW, 
Washington, DC 20052, USA
}
\author[0000-0002-8548-482X]{J. Hare}
\affiliation{NASA Goddard Space Flight Center, Greenbelt, MD 20771, USA 
}



\begin{abstract}
Previous observations of the middle-aged $\gamma$-ray, X-ray, and radio pulsar B1055--52 indicated some peculiarities, such as
a suspected changing of the X-ray flux and spectral parameters, a  
large excess of the alleged thermal component of the ultraviolet (UV) spectrum over the Rayleigh-Jeans extension of the X-ray thermal spectrum, and a possible double break in the nonthermal spectral component between the optical and X-ray bands.
We observed PSR B1055--52 with the {\sl XMM-Newton} observatory in X-rays and the {\sl Hubble Space Telescope} in near-infrared (NIR).
The analysis of the {\sl XMM-Newton} observations does not support the notion of long-term changes in the X-ray flux and broad-band X-ray spectrum of the pulsar. Using an observing mode less affected by background noise than the previous XMM-Newton observations, we constrain the power-law (PL) spectral index as  $\alpha_X=-0.57^{+0.26}_{-0.25} $ ($F_{\nu} \propto \nu^{\alpha}$) in the energy band 3--10\,keV. 
From the NIR-optical data we obtain a PL slope 
$\alpha_O= -0.24 \pm 0.10$ for the color index $E(B-V)=0.03$\,mag.
The slopes and fluxes of the NIR-optical and X-ray nonthermal spectra suggest that the NIR through X-ray emission can be described by the same PL and is generated by the same mechanism, unlike the pulsar's $\gamma$-ray emission.
The excess of the UV thermal component over the extension of the X-ray thermal component became smaller but did not disappear, indicating a non-uniformity of the bulk surface temperature.
The NIR data also enable us to accurately measure the proper motion with values  $\mu_\alpha =47.5\pm 0.7\,{\rm mas\,yr}^{-1}$ and $\mu_\delta = -8.7 \pm 0.7 \,{\rm mas\,yr}^{-1}$.

\end{abstract}


\section{Introduction} 
\label{sec:intro}
Radiation of rotation powered pulsars includes a nonthermal
component, emitted by relativistic
particles in the pulsar magnetosphere,
and a thermal component, emitted from the neutron star (NS) surface.
Multiwavelength 
observations
enable one to understand
the mechanisms of particle acceleration and emission, the properties
of the magnetosphere and the NS surface, and thermal evolution of NSs.
Comparison of UV-optical-IR (UVOIR)
spectra and pulse profiles with the X-ray ones, for example,
are crucial to measure the
NS surface temperatures,
probe the temperature nonuniformity,
estimate the NS radii,
and measure the luminosities and spectral slopes
of the magnetospheric (synchrotron) emission
(e.g., \citealt{Kaplan2011,Mignani2011,Kargaltsev2007, Pavlov2002})\\

The thermal and nonthermal components evolve differently with pulsar age (e.g., \citealt{Pavlov2002}; for a recent statistical compilation see \citealt{Chang2023}). Although the NS surface is expected to be the hottest in very young pulsars, their thermal emission is buried under the powerful magnetospheric emission. At ages of  $\sim 100$ kyr the magnetospheric emission becomes fainter, and the X-ray thermal emission emerges, with a maximum flux in the extreme UV to soft X-ray range. This thermal emission consists of a {\em cold component} associated with the residual heat of the cooling NS,
and a {\em hot component} from a fraction of the NS surface presumably heated by relativistic particles precipitating from the pulsar magnetosphere. When the pulsar becomes older than $\sim 1$ Myr, the NS becomes too cold to emit in X-rays, so the cold component is only detectable in UV-optical, but the hot thermal component can still be seen in soft X-rays, together with the (faint) magnetospheric component. Thus,
multiwavelength observations of {\em middle-aged} ($\tau \sim 0.1$--1\,Myr) pulsars allow one to study all the components of the NS's electromagnetic radiation, which is 
particularly important for understanding 
NS physics and evolution.\\

Well-studied middle-aged NSs are the nearby PSRs B0656+14, J0633+1746, 
and B1055--52 (B0656, Geminga, and B1055 hereafter), with periods $P=385$, 237 and 197 ms, spin-down ages $\tau = P/(2\dot{P}) = 111$, 342 and 535 kyr, 
and spin-down powers 
$\dot{E} = 3.8$, 3.2 and 3.0 $\times 10^{34}$\,erg s$^{-1}$,
and surface magnetic fields 
$B = 4.7$, 1.6 and 1.1 $\times 10^{12}$\,G.
While the distances to B0656 and Geminga, $d=288$ and 250\,pc 
respectively, have been found from parallax measurements \citep{Brisken2003,Faherty2007}, the distance to B1055 remains uncertain; the Galactic free electron density models by \citet{Cordes2002} and \citet{Yao2017} give 714 and 93 pc, respectively, for the B1055's dispersion measure ${\rm DM} = 29.7$ cm$^{-3}$ pc$^{-1}$ \citep{Petroff2013}.
All three are $\gamma$-ray and X-ray pulsars, with maximum energy emitted in GeV $\gamma$-rays. Thermal emission dominates their soft X-rays spectra at $E\lesssim 2$ keV.
Because of the similarity of their X-ray and spin-down properties, they were 
dubbed {\em the Three Musketeers} by \citet{Becker1997}. All three have been detected in the UV-optical, but only in the former two optical-UV pulsations were found and (N)IR emission detected. 
While B0656+14 and Geminga have not shown any variations in their emission, a $\gtrsim 30$\% change of the B1055 X-ray flux has been reported by \citet{Posselt2015}.
Phase-resolved spectroscopy of {\sl Chandra} and XMM-\emph{Newton} data from 2000 \citep{Pavlov2002,deLuca2005}
showed that,
similar to B0656 and Geminga, B1055's X-ray emission
consists of 
a cold thermal component
with a
temperature $T_{X,C} = 0.8$\,MK, 
emitted
from a
substantial fraction
of the NS surface, a hot thermal component ($T_{X,H} = 1.8$\,MK) from a much smaller area, and a
magnetospheric component with a power-law (PL) spectrum,
$F_{\nu}\propto\nu^{\alpha_{\rm X}}$ (spectral index $\alpha_{\rm X}\simeq -0.7$).\\

Comparing the 2000 XMM-\emph{Newton} and 2012 {\sl Chandra} imaging observations, \citet{Posselt2015} reported a surprisingly large difference of fluxes (flux decrease of $47$\% if all spectral parameters are independently fit) and spectral parameters (flux decrease of $29$\% and hotter thermal emission from a smaller emitting area in 2012 if some spectral fit parameters were tied for the {\sl Chandra} and XMM-\emph{Newton} data). 
Uncertainties due to cross-calibration errors between the XMM-\emph{Newton}'s EPIC and {\sl Chandra}'s ACIS instruments are expected to be much smaller. For instance, $\sim 2$\% in flux differences between  
EPIC-pn (PN in the following) and ACIS were reported at energies $0.5-2$\,keV for data until 2009 in the cross-calibration study of 11 galaxy clusters by \citet{Nevalainen2010}. For one X-ray thermal isolated neutron star RBS\,1223, \citet{Haberl2003} reported a 10\% flux difference between EPIC and ACIS\footnote{For other reference studies, see section 4.2 in \citet{Posselt2015}.}. 
Therefore, the X-ray difference found by \citet{Posselt2015} raised the interesting possibility that 
B1055 may exhibit long-term spectral variability. Except for magnetars, large variability at high energies is unusual for middle-aged isolated NSs. Rare examples are the $\gamma$-ray-ray pulsar PSR J2021+4026 (sudden 20\% flux decrease above 100\,MeV; \citealt{Allafort2013}) and the X-ray thermal isolated NS RX\,J0720.4$-$3125 ($\sim 10$\% flux change in the soft X-ray band $0.12-1.0$\,keV and significant variation of the spectral parameters; e.g., \citealt{Hohle2012}). Such changes can be explained by glitches, a re-arrangement of the
magnetosphere, or impacts of circumstellar material on the NS surface \citep{Zhao2017,Hohle2012,Kerkwijk2007}.\\

Optical observations with ground-based telescopes failed to detect B1055
because of the presence of a bright ($V =14.6$)
F-type star
$\approx4''$ from the pulsar position
\citep{Bignami1988}.
A faint ($U\approx 25$) candidate counterpart was detected 
by \cite{Mignani1997} in the
F342W band (3060--3760\,\AA)
with the {\sl{Hubble Space Telescope}} (\emph{HST}) Faint Object Camera (FOC).
The identification was confirmed by the next {\sl HST} observation of 2008 (\citealt{Mignani2010}; \citetalias{Mignani2010} hereafter). The pulsar was detected in the optical bands F555W (4830--6060\,\AA) and F702W (6230--7610\,\AA) with the Wide Field Planetary Camera 2 (WFPC2), and in the far-UV (FUV) band 1400--1700 \AA\ (longpass filer F140LP) with the Solar Blind Channel (SBC) of the Advanced Camera for Surveys (ACS).
From the comparison with the previous HST image, \citetalias{Mignani2010} provided the first estimate of the pulsar's proper motion:
 $\mu_\alpha =42\pm 5\,{\rm mas\,yr}^{-1}$ 
 and $\mu_\delta = -3\pm5\,{\rm mas\,yr}^{-1}$.
Assuming that, similar to B0656 and Geminga, the 
UVOIR spectrum of B1055 consists of nonthermal and thermal components, dominating in NIR-optical and (F)UV, respectively, \citetalias{Mignani2010} fit the four flux density 
points with a two-component 
PL + Rayleigh-Jeans (RJ) model. 
They found the optical  PL slope $\alpha_O =-1.05\pm 0.34$, 
PL normalization $f_O=112\pm 3$ nJy (at $\nu_0= 1\times 10^{15}$ Hz),  and `RJ parameter' $R_O^2T_O =
(111\pm18) d_{350}^2$ km$^2$\,MK,
for an assumed color index $E(B-V) =0.07$. Here $R_O$ is the radius of an  equivalent emitting sphere, $T_O$ is the brightness temperature (both as seen by a distant observer), and 
$d_{350}$ is the distance in units of 350 pc. The PL component, which can be interpreted as magnetospheric radiation, dominates  in the optical ($\lambda\gtrsim 3000$\,\AA), while the RJ component dominates in the FUV and is likely emitted from the NS surface.\\

Comparing the optical-UV spectral parameters with the X-ray spectrum of B1055--52, 
\citetalias{Mignani2010} found that the optical PL component is significantly steeper than the X-ray one,
which could be due to different emission mechanisms in the optical and X-ray ranges. Also, the UV-optical RJ component showed a factor of 4 excess over the extrapolation of the X-ray thermal component into the UV-optical range, which hinted that the RJ component is  emitted from a larger, colder area. However, because of the very short frequency range covered by the optical filters, the uncertainties of the optical PL slope were rather large, which in turn led to large uncertainties of the RJ parameter (or brightness temperature).
Considering also the $\gamma$-ray spectral range, there remained the question whether or not the nonthermal X-ray component can have the same slope as the $\gamma$-ray component detected by the Fermi observatory.\\ 

\begin{deluxetable*}{lllccccc}[t!]
\tablecolumns{8}
\tablecaption{XMM-Newton observations} \label{XMMobs}
\tablewidth{0pt}
\tablehead{
\colhead{Observation ID} & \colhead{MJD} &  \colhead{Detector (Mode, Filter)}  & \colhead{Exp time} & \colhead{NCR (0.3--10\,keV)} & \colhead{SF} & \colhead{NCR (3--10\,keV)} & \colhead{SF} \\
\colhead{ } & \colhead{ } &  \colhead{ }  & \colhead{ks} & \colhead{cts ks$^{-1}$} & \colhead{\%} & \colhead{cts ks$^{-1}$} & \colhead{\%} 
}
\startdata
0842820101 & 58654 & pn (SW, Thin)     & 51.7 &  $764.2\pm 3.9$ & 97 &  $3.90 \pm 0.48$  & 38 \\
           &       & MOS1 (FF, Medium) & 75.5 &  $124.7\pm 1.3$ & 96 &  $0.86 \pm 0.20$  & 30 \\
           &       & MOS2 (FF, Medium) & 75.0 &  $114.4\pm 1.3$ & 96 &  $0.81 \pm 0.21$  & 28 \\
0842820201 & 58673 & pn (SW, Thin)     & 54.4 &  $738.2\pm 3.7$ & 97 &  $4.17 \pm 0.44$  & 41 \\
           &       & MOS1 (FF, Medium) & 77.7 &  $120.4\pm 1.3$ & 96 &  $0.99 \pm 0.21$  & 32 \\
           &       & MOS2 (FF, Medium) & 77.7 &  $121.3\pm 1.3$ & 96 &  $1.28 \pm 0.21$  & 42 \\
0113050101 & 51892 & pn (Timing, Medium)&19.3 &  $444.7\pm 6.2$ & 83 &  $2.13 \pm 1.80$  & 10 \\
           &       & MOS1 (FF, Medium) & 21.0 &  $141.7\pm 2.6$ & 98 &  $1.01 \pm 0.36$  & 52 \\
           &       & MOS2 (FF, Medium) & 21.0 &  $147.7\pm 2.7$ & 99 &  $1.12 \pm 0.32$  & 65\\
0113050201 & 51893 & pn (Timing, Medium)&51.1 &  $435.8\pm 3.8$ & 82 &  $3.56 \pm 1.10$  & 17\\
           &       & MOS1 (FF, Medium) & 53.4 &  $147.5\pm 1.7$ & 99 &  $1.66 \pm 0.24$  & 68 \\
           &       & MOS2 (FF, Medium) & 53.4 &  $148.9\pm 1.7$ & 98 &  $1.28 \pm 0.22$  & 63\\
\enddata
\tablecomments{The modified Julian dates (MJDs) correspond to the start date of the observation in the {\sl XMM-Newton} archive. 
The effective exposures times for each EPIC detector are reported after screening for background flares and dead time correction. The observing modes SW, FF correspond to Small Window mode and Full Frame mode, respectively.
The net (i.e., background-subtracted) count rates, NCR, and the source count fraction, SF,
are listed for the two energy ranges for which the fits of 12 spectra were carried out. For individual detectors, the actual energy range may be smaller, e.g., the EPIC-pn timing mode data only includes events above 0.4\,keV.
}
\end{deluxetable*}

We present here new {\sl XMM-Newton} observations of B1055 that were obtained to investigate the discrepancy of the X-ray spectral parameters and fluxes of the 2000 {\sl XMM-Newton} and 2012 \emph{Chandra} as well as to improve the count statistiscs at $>3$\,keV for a tighter constrain of the PL.
We compare this PL with the 
up-to-date \emph{Fermi} $\gamma$-ray spectrum.
We also present our additional {\sl HST} observations in two 
NIR bands. These data are used to measure the optical PL slope 
and the RJ parameter 
more accurately and to compare the optical and X-ray spectra more reliably. 
In addition, the new {\sl HST} observation also allowed us to measure the pulsar's proper motion more accurately than it has been possible so far. 

\section{Observations \& Data Analysis} 
\label{sec:obsana}
\subsection{{\sl XMM-Newton} observations}
\label{sec:obsxmm}
New {\sl XMM-Newton} observations were carried out on 2019 June 20 and 2019 July 9 for 85\,ks and 81\,ks, respectively (Observation IDs 0842820101 and 0842820201; PI Posselt). The pn detector of the EPIC (European Photon Imaging
Camera) instrument was 
operated in Small Window (SW) imaging mode with Thin filter, the MOS1 and MOS2 cameras in Full Frame (FF) imaging mode with the Medium filter. 
We also re-analysed the {\sl XMM-Newton} observations from 2000 (PN in timing mode, MOS1 and MOS2 cameras in FF mode, all with the Medium filter). These data were first presented by \citet{deLuca2005}.
The parameters of the observations are summarised in Table~\ref{XMMobs}. 
The spectral extraction for the data from 2000 was done in the same way as already described in detail in \citet{Posselt2015}.
This time, however, an updated version 18.0 of the {\sl XMM-Newton} \emph{Science Analysis Software}, SAS \citep{Gabriel2004} was used for all our data reduction. We note that these early {\sl XMM-Newton} PN observations do not have associated offset maps. Because the soft photon background noise can therefore influence the data, our extracted PN spectra from 2000 only cover energies larger than 400\,eV. The 2019 PN SW data were filtered for the soft X-ray noise using the SAS task \texttt{epnoise}.
We also filtered the data for background flares in all EPIC instruments. 
For the imaging data, 
we excluded two (weak) X-ray sources at separations of $30\arcsec$ and $39\arcsec$
from B1055, best seen in previous {\sl Chandra} observations (Figure~1 of \citealt{Posselt2015}). For this we used circular regions with radii of $15\arcsec$.
The source extraction regions centered on the pulsar had a radius of $35\arcsec$ for PN, and $45\arcsec$ for the MOS cameras. Background regions were chosen in source-free regions on the same detector chip. 
We extracted spectra using events with pattern 0--12 for MOS, but only single and
double photon events (pattern 0-4) for PN.
We tested different binnings of the spectra, employing the \texttt{specgroup}-task using an oversampling factor of 3 with similar spectral fit results. Here, we report all values for data binned to have at least 15 counts per bin. This binning allowed us to include more high-energy channels.\\ 

Following the procedure outlined in \citet{Jethwa2015}, we checked whether our PN SW and MOS FF imaging mode
observations are subject to significant pile-up, i.e., the counting of several X-ray photons as one. Even using our larger extraction regions\footnote{A $30\arcsec$ radius circle is used by \citet{Jethwa2015}}, all our count rates or  incoming number of photons per frame stay below the conservative limits estimated to cause less than 1\% spectral distortion due to pile-up. For the 0842820101 PN spectrum, for example, using deadtime-corrected exposure time of 51.7\,ks, we estimated that there are $\sim 0.01$\, photons per frame, corresponding to spectral distortions of $<0.3$\,\% and flux losses $<2$\,\% according to Figure~5 by \citet{Jethwa2015}. 
Thus, we conclude that pile-up is not significant for any of the 
{\sl XMM-Newton} observations of B1055.\\ 

In this paper, we only present spectral analysis with the goal to test for 
the previously suspected {X-ray} flux and spectrum variations within 19 years and derive tighter spectral constraints for the multiwavelength spectral energy distribution of this pulsar. The timing and phase-resolved spectral analysis of the X-ray 
data together with the accompanying radio observations will be presented in another paper (Vahdat et al., \emph{in preparation}).
 As previously demonstrated \citep{Posselt2015,deLuca2005,Pavlov2002}, NS atmosphere (plus nonthermal component) models
required unrealistically large radius-to-distance ratios (even 
with account for distance uncertainty)
and resulted in worse fits than a spectral model consisting of two thermal blackbody (BB) components and a power law (PL), which describe the X-ray data well. 
Employing this model, we first investigate the best-fit values for the new and old XMM-{\sl Newton} data separately to ensure agreement. Then,  
we carry out simultaneous spectral fits of the phase-integrated data for all the individual twelve {\sl XMM-Newton} spectra using XSPEC (version 12.10, \citealt{Arnaud1996}) with the three-component spectral model and the T\"ubingen-Boulder ISM absorption model (\texttt{tbabs}) with the solar abundance table from \citet{Wilms2000} and cross-sections by \citet{Verner1996}.\\

The new PN SW imaging data are less influenced by background noise at high energies, 
which results in a substantial increase in count statistics above 3\,keV (see Table~\ref{XMMobs} for the count rates).  
However, even this improved count statistic is still poor. Because of this, and because of correlations between the spectral components, in particular the PL and the hot BB, the inclusion of lower energies could lead to a biased 
photon index.
Therefore, we employ a two-step procedure. First, we fit a simple PL in the energy range 3--10\,keV,
assuming a fixed hydrogen column density, $N_H =1.4\times 10^{20}$\,cm$^{-2}$ 
\footnote{At the considered energies and at such small column densities the PL fit result is insensitive to the exact $N_H$ value.}.
The derived photon index, $\Gamma = 1.57^{+0.26}_{-0.25}$, is then fixed for the 2BB+PL fit in the 0.3--10\,keV energy range while $N_H$ is allowed to vary. Since a fixed  $\Gamma$ parameter reduces the uncertainty range of the BB parameters, we
obtain the latter by re-fitting the thermal components  with $\Gamma$ being fixed at its  lower and upper 1$\sigma$ bounds  (1.32 and 1.83) and then combining the uncertainties of the thermal components from both fits.
The results for the twelve spectra (i.e. assuming no variability between the two epochs) are reported in Table~\ref{XMMresult}. 
For completeness, we show in Appendix~\ref{APLfit} the dependency of $\Gamma$ on the lower energy boundary. 
The fit from a single 2BB+PL-fit with free PL parameter is formally as good as the fit presented in Table~\ref{XMMresult}. Uncertainty ranges from the two fit methods overlap\footnote{This can be ssen in the comparison of the model plots in Section~\ref{XrayGamma} and Appendix~\ref{APLfit}.}. In the following we use the results from the two-step fit whose photon index we regard as less influenced by the thermal model component.

\begin{deluxetable}{lc}[t]
\tablecolumns{2}
\tablecaption{X-ray spectral fit results} \label{XMMresult}
\tablewidth{0pt}
\tablehead{
\colhead{Parameter} & \colhead{value}
}
\startdata
\multicolumn{2}{l}{PL fit, 3-10\,keV}\\
\hline
$N_{\rm H}$ (10$^{19}$ cm$^{-2})$ &       14 (fixed)                                \\
$\Gamma$ & $1.57^{+0.26}_{-0.25}$                                        \\
$K_{\rm X,PL}$ & $1.20^{+0.57}_{-0.39}$                                       \\
$\chi^{2}$/dof (dof) &   1.1 (229)                                     \\
$\mathcal{F}_{\rm pn}$ (2000/2019)&          1 / 1 (fixed) \\
$\mathcal{F}_{\rm MOS1}$ (2000/2019) &    $1.2\pm0.2$ / $0.8 \pm 0.1$ \\
$\mathcal{F}_{\rm MOS2}$ (2000/2019) &    $1.0\pm0.2$ / $0.8 \pm 0.1$  \\
$F^{\rm PL}_{\rm unabs, 3-10 keV}$ (10$^{-14}$\,erg\,cm$^{-2}$\,s$^{-1}$) &  $4.8 \pm 0.5$       \\
\hline
\multicolumn{2}{l}{2BB+PL fit, 0.3--10\,keV}\\
\hline
$N_{\rm H}$ (10$^{19}$ cm$^{-2})$ &    $14^{+3}_{-2}$  \\
$kT_{\rm X,C}$ (eV) &                  $70.2^{+1.6}_{-1.7}$\tablenotemark{a} \\
$R_{\rm X,C}$ (km) &                   $4.7^{+0.5}_{-0.4}$  \\
$kT_{\rm X,H}$ (eV) &                  $165\pm 12$\tablenotemark{a}\\
$R_{\rm X,H}$ (m) &                    $180^{+50}_{-30}$ \\
$\Gamma$ &    1.57 (fixed)\tablenotemark{b}          \\
$K_{\rm X,PL}$ &               $1.23 \pm {0.04}$     \\
$\chi^{2}$/dof (dof) &                 1.2    (825)                               \\
$\mathcal{F}_{\rm pn}$ (2000/2019) &      1 / 1 (fixed)  \\
$\mathcal{F}_{\rm MOS1}$ (2000/2019) &  $1.06\pm 0.01$ / $1.052 \pm 0.009$ \\
$\mathcal{F}_{\rm MOS2}$ (2000/2019) &  $1.04\pm 0.01$ / $1.063 \pm 0.009$ \\
$F_{\rm abs, 0.3-10 keV}$ (10$^{-14}$\,erg\,cm$^{-2}$\,s$^{-1}$) &  $167.6 \pm 0.8$     \\
$F^{\rm PL}_{\rm unabs, 0.5-8 keV}$ (10$^{-14}$\,erg\,cm$^{-2}$\,s$^{-1}$) &  $7.8 \pm 0.3$       \\
$L^{\rm PL}_{\rm 0.5-8 keV}$ ($10^{30}$ erg s$^{-1})$ & $1.14 \pm 0.04$ \\
$L_{\rm bol, X,C}$ (10$^{31}$ erg s$^{-1}$) &     6.9 \\
$L_{\rm bol, X,H}$ (10$^{30}$ erg s$^{-1}$) &     3.1 \\
\hline
\enddata
\tablecomments{The fit is carried out for 12 spectra simultaneously. The PL luminosity and radius calculations were done using a distance value of 350\,pc and the exact best-fit parameters before rounding.
The following best-fit parameters are given: absorbing hydrogen column density $N_{\rm H}$, BB temperature $kT$, normalization $K_{\rm X,PL}$ of the PL component in units of 10$^{-5}$\,photons\,keV$^{-1}$\,cm$^{-2}$\,s$^{-1}$ at 1 keV, radius $R$ for the equivalent sphere of the BB emission, photon index $\Gamma$ of the PL component, fluxes $F$, bolometric luminosities of the BBs, the luminosity of the PL component in the energy range $0.5-8$\,keV, reduced $\chi^2$, the number of degrees of freedom, and the calibration factors $\mathcal{F}$ for pn, MOS1 and MOS2.
Uncertainties indicate the 68\% confidence level. The uncertainties of the BB components encompass those obtained for $\Gamma$ fixed at its 68\% confidence levels. The PL luminosity uncertainties do not include the distance uncertainties.}
\tablenotetext{a}{Or $T_{\rm X,C} =0.815^{+0.019}_{-0.020}$\,MK,
$T_{\rm X,H}=1.91\pm 0.14$\,MK.}
\tablenotetext{b}{{For the uncertainties of the thermal components, we also consider the fits where $\Gamma$ is fixed at 1.32 and 1.83.}}
\end{deluxetable}

\subsection{{\sl HST} observations}
\label{sec:obshubble}
\begin{figure}[t]
\includegraphics[width=8.5cm]{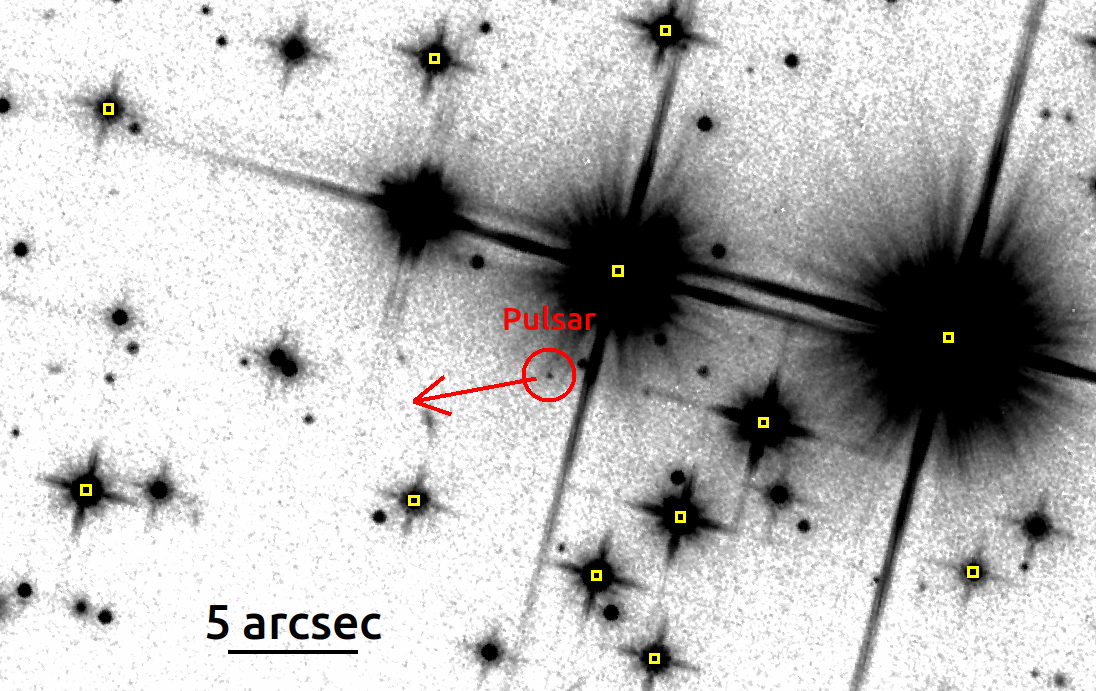}
\caption{The pulsar\,B1055$-$52 and its environment as detected in our HST F110W observation. North is up, east to the left. The yellow squares indicate \emph{Gaia} sources, corrected for their proper motion (if known) to the HST observing epoch. The red circle with radius 1\,arcsec indicates the pulsar location. The red arrow shows the direction of the pulsar proper motion. \label{fig:overview}}
\end{figure}

The pulsar was observed 
with the \emph{HST} Wide Field Camera 3 (WFC3; \citealt{Kimble2008}) 
IR channel on 2019 April 29 (MJD 58602) 
in one \emph{HST} orbit (program \#15676; PI Posselt).
The IR Channel is a HgCdTe detector with a $136\arcsec\times 123\arcsec$ field-of-view  and a pixel scale of 0\farcs13 pix$^{-1}$.
The observations were carried out for 648\,s and 1998\,s with the F110W (wide YJ) and F160W (WFC3 H) filters, with pivot wavelengths of $1.153\,\mu$m  and $1.537\,\mu$m
and widths of 0.443\,$\mu$m  and 0.268\,$\mu$m, respectively. 
We used the WIDE-3POS-TARG dither pattern (optimizes the subsampling of the pixels\footnote{{See Instrument Science Report} ISR 2016-14, Supplemental Dither Patterns for WFC3/IR, J.\ Anderson;  \url{https://www.stsci.edu/hst/instrumentation/wfc3/documentation/instrument-science-reports-isrs}}) for our 3 individual exposures per band. 
We employed the `orient' constraint, restricting the orientation of the instrument on the sky, in order to ensure that the pulsar position is  not affected by diffraction spikes from the bright stars close to it (see Figure~\ref{fig:overview}).
Exposures were taken in MULTIACCUM mode with the rapid-log-linear  timing sequences STEP25 (F110W) and STEP100 (F160W) increasing the dynamic range of the final images. 
We stacked and processed our images using
PyRAF and AstroDrizzle (version 2.1.8) of the DrizzlePAC software \citep{Gonzaga2012,Fruchter2010}, with the inverse variance map (IVM) weighting scheme for the final combination of the data \citep{Gonzaga2012}. We experimented with the drizzle parameters for a balance of spatial resolution and sampling noise. We chose a final pixel scale of $32$\,mas\,pixel$^{-1}$ 
and a pixel fraction of 0.9 (the fractional area of overlap of a pixel that is considered in the drizzle process; for details see \citealt{Gonzaga2012}) 
for both bands.  
An overview of the larger region around the pulsar is shown for band F110W in Figure~\ref{fig:overview}, while zoom-ins onto a small region around the pulsar in both bands can be found in the Appendix.  

\subsubsection{Astrometry}
\label{sec:astrometry}
\begin{figure*}[t]
\includegraphics[width=8.7cm]{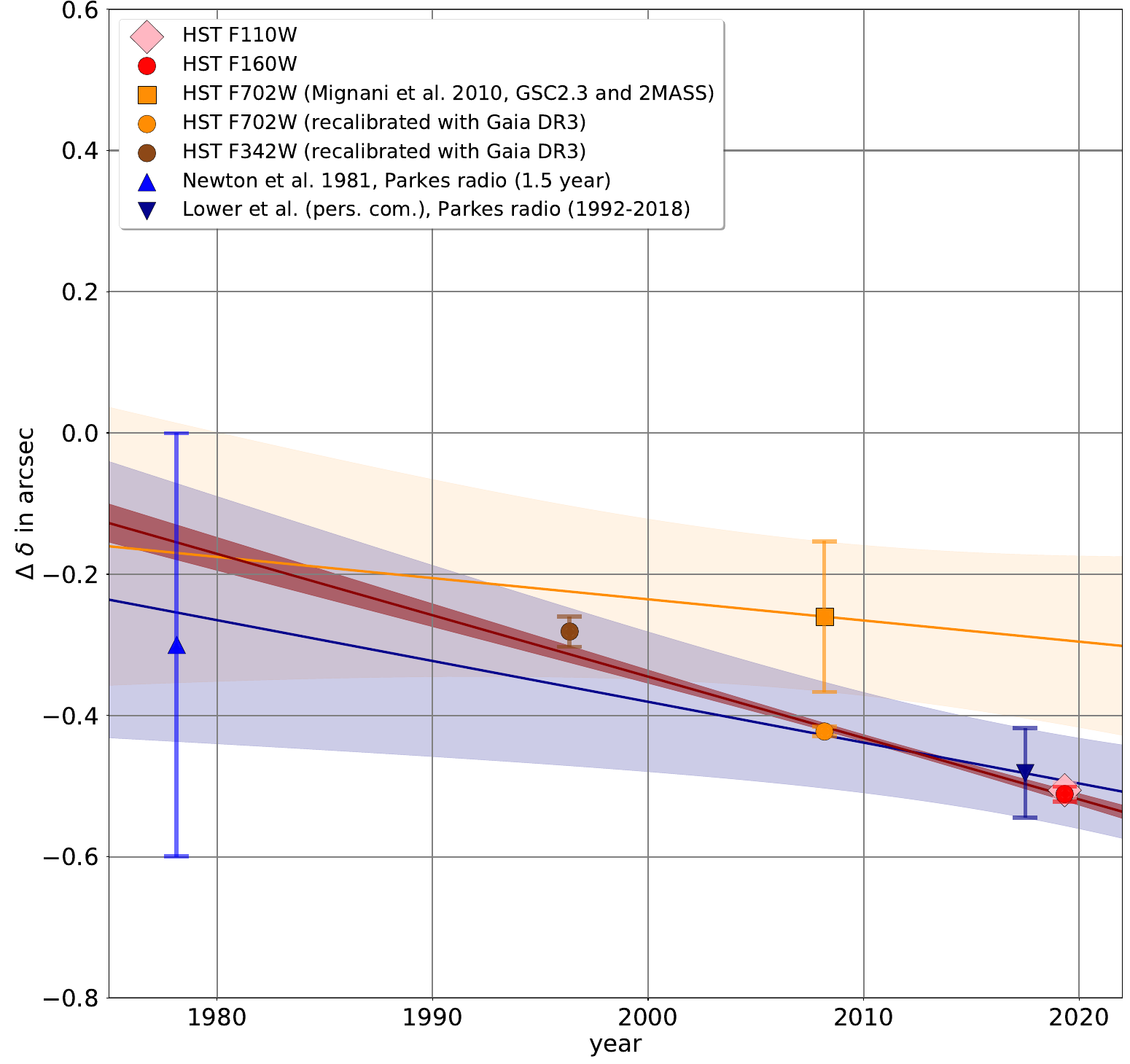}
\includegraphics[width=8.7cm]{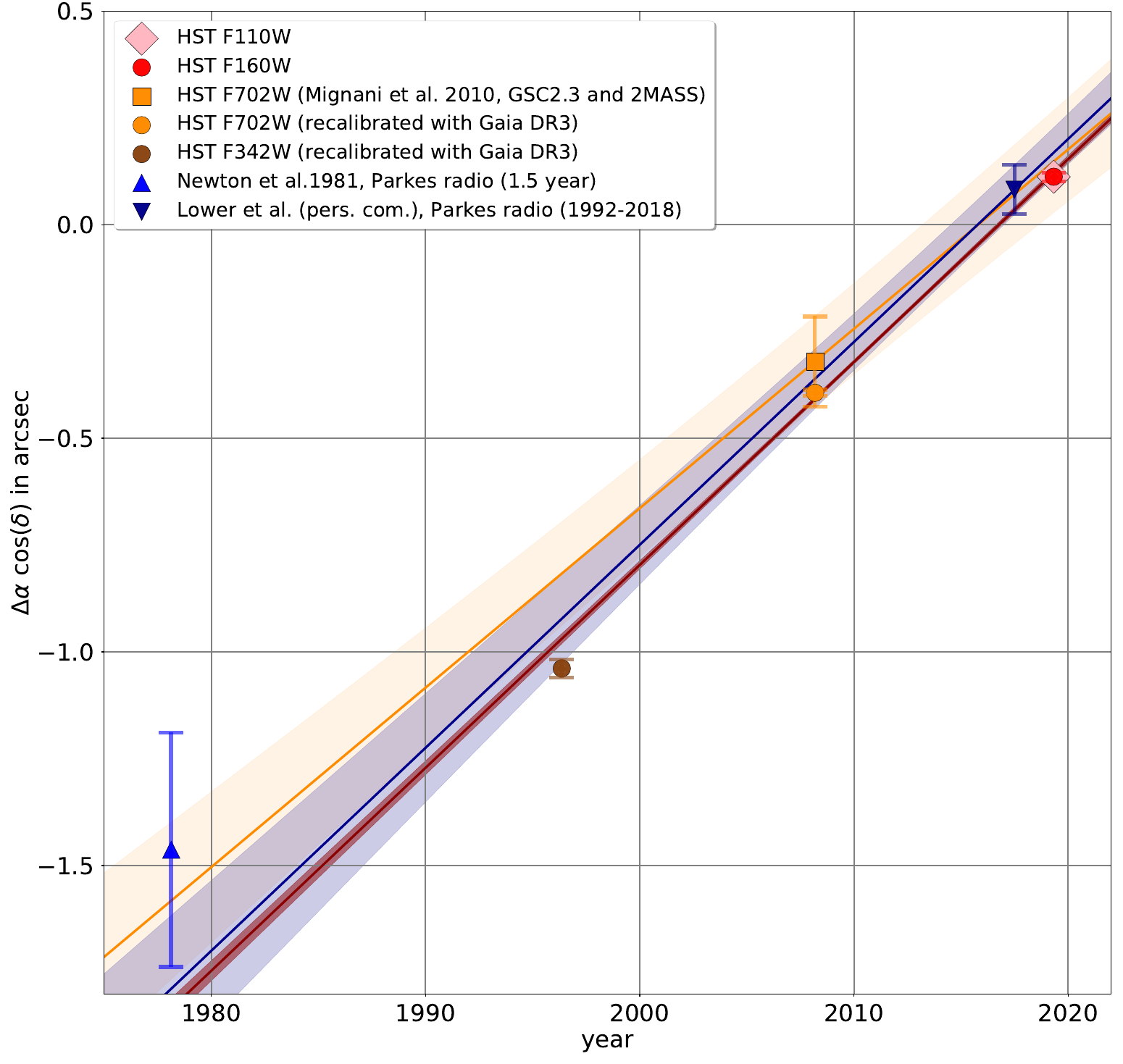}
\caption{Positions and proper motion estimates for PSR\,B1055$-$52. The left and right plots show the declination and right ascension coordinates, with the zero point at 
R.A.$= 10^{\rm h} 57^{\rm m} 59\fs0$ and decl.$=-52\arcdeg 26\arcmin 56\farcs0$. The yellow, blue, and red 
 lines and shaded regions mark the proper motions estimates and their $1\sigma$ uncertainty regions according to \citet{Mignani2010}, Lower et al.\ (private communication; result obtained from 26 years of available Parkes data), and this work, respectively.}
\label{fig:pm}
\end{figure*}

\begin{deluxetable*}{lcccccllc}[t!]
\tablecolumns{8}
\tablecaption{Pulsar astrometry} \label{positions}
\tablewidth{0pt}
\tablehead{
\colhead{Detector \& band} & \colhead{epoch} & \colhead{pixel scale} & \colhead{$\#$ stars} & \colhead{$\delta^{\rm radial}_{\rm sys}$} & \colhead{$\delta^{\rm radial}_{\rm cent}$} & \colhead{R.A.} &  \colhead{decl.}   & \colhead{$\delta^{1D}$}\\  
\colhead{} & \colhead{} & \colhead{mas/pix} & \colhead{} & \colhead{mas} & \colhead{mas} & \colhead{$-10^{\rm h}$ 57$^{\rm m}$} &  \colhead{ $+ 52^\circ$ $26\arcmin$} &\colhead{mas}     
}
\startdata
FOC 342W & 1996.4 &  14.4 & 1 &  29 &  7 & $58\fs8864$ & $-56\farcs281$ & 21 \\
\hline
WFPC2 F702W & 2008.2 & 49.9 & 14 & 8 &  5  & $58\fs9570$ & $-56\farcs423$ & 7 \\ 
\hline
WFC3 F110W & 2019.3 & 32.1 & 81 &  14  &   6 & $59\fs0123$ & $-56\farcs506$ & 11\\ 
WFC3 F160W & 2019.3 & 32.1 & 81 & 13 &  6 & $59\fs0123$ & $-56\farcs511$ & 10\\ 
WFC3 NIR (weighted) & 2019.3 &  &  & &  & $59\fs0123$ & $-56\farcs509$ & 8\\
\enddata
\tablecomments{
The numbers of field stars used for the astrometry is shown in the fourth column.
The fit to the \emph{Gaia} DR3 counterparts of these field stars resulted in the systematic radial uncertainty $\delta^{\rm radial}_{\rm sys}$, while $\delta^{\rm radial}_{\rm cent}$ indicates the radial centroiding uncertainty. The last three columns list the pulsar coordinates and the total uncertainties in \emph{each} coordinate direction.}
\end{deluxetable*}

The accurate astrometry of the \emph{Gaia} data release 3, DR3 
\citep{GaiaDR3_2022,Gaiaedr3_2021}
allows us not only to calibrate the astrometry in our recent {\sl HST} WFC3 NIR data, but also to substantially improve the astrometry of the previous {\sl HST} observations from 1996 and 2008 (FOC and WFPC2 instuments). Since this can enable a more accurate proper motion measurement, we carry out the astrometry calibration for these previous {\sl HST} epochs, too.
For the two earlier epochs, we used processed data obtained from the Barbara A. Mikulski Archive for Space Telescopes (MAST)\footnote{See \url{http://archive.stsci.edu/hst/}}. The data can be accessed at \dataset[10.17909/2td7-8m31]{https://doi.org/10.17909/2td7-8m31}.
We applied the catalog proper motions to calculate the positions of the selected \emph{Gaia} DR3 stars at the three {\sl HST} observing epochs.
For the WFPC2 and WFC3 data, we used the Graphical Astronomy and Image Analysis Tool \citep{Currie2014} to obtain a fit of the image astrometry with the proper-motion corrected \emph{Gaia} DR3 catalog positions. The root mean square of the respective fit residuals is considered the systematic astrometric error (with respect to the \emph{Gaia} DR3 reference frame). They are listed in Table~\ref{positions}. 
In the 1996 FOC $\sim 7$\arcsec $\times 7$\arcsec\,images,
there is only one field star (star A in \citet{Mignani1997,Mignani2010}, marked in Figure~\ref{fig:overview}; star A is also listed in \emph{Gaia} DR3, with proper motion of 
$\mu_\alpha =-7.848\pm 0.014$\,mas\,yr$^{-1}$, $\mu_\delta = 1.676 \pm 0.014$\,mas\,yr$^{-1}$; \citealt{GaiaDR3_2022}), which is bright and saturated. We used separately the three individual FOC exposures, diffraction spikes, and the roundish surface brightness distribution to determine a systematic error of star A's position.  
There is a remaining unknown systematic error since image distortions or rotations could not be constrained with only one reference position. 

Considering the small absolute astrometric error of the NIR observations (see Table~\ref{positions}), there is no doubt that the pulsar is detected in F110W and F160W close to the position expected according to the proper motion estimate by \citet{Mignani2010}. In order to check for the presence of extended emission around the pulsar, we compared the FWHMs of the pulsar's image and those of the field stars.
Using about 20 
stars in the field of view, we find a median FWHM of 151\,mas (145\,mas) with standard deviation of 43\,mas (53\,mas) for the F110W (F160W) bands. For the pulsar, the measured FWHM is about 180\,mas in both NIR bands, well within the uncertainty range of our FWHM measurements.
It should be noted that the pulsar is located in an inhomogeneous noisy background region because it is close to several bright objects and their diffraction spikes, see Figure~\ref{fig:overview}.
Comparing the two bands, we noticed noise features at different position angles that likely gave rise to the slighly larger FWHM measurements for the pulsar in comparison to the median FWHMs. Thus, we do not see an indication of an extended source. \\

For the WFPC2 and WFC3 data, we derived centroid errors for the pulsar positions using FWHM values from 
2D Gaussian fits, and the peak signal-to-noise ratio of the detections following 
Equation (1) in \citet{Reid1988}. The resulting centroid errors are  between 0.1--0.2\,pix (3--6\,mas).
For the FOC data, we estimated the centroid error to be smaller than 0.5\,pix (7\,mas).
These combined (centroid and systematic) radial uncertainties, $\delta_r$, were converted into uncertainties in R.A. and declination assuming 63\% of the source counts are within a circular area with radius equal $\delta_r$.
The pulsar positions and their uncertainties in the individual HST observations and the respective observational parameters are summarized in Table~\ref{positions}.\\

From the three HST positions and their uncertainties
listed in Table~\ref{positions}, we calculated the proper motion at the (uncertainty-weighted) middle epoch, 
2010.96,
\begin{equation}
\label{pm}
 \mu_\alpha =47.5\pm 0.7\,{\rm mas\,yr}^{-1}, \quad \mu_\delta = -8.7 \pm 0.7 \,{\rm mas\,yr}^{-1}.
\end{equation}

Since we do not know the systematic error of the FOC astrometry, we checked for the influence of the FOC measurement on the result.
Increasing its systematic error to 100\,mas 
gives a proper motion estimate at the middle epoch 
2012.83 of  
\begin{equation}
\label{pmbigFOCerror}
 \mu_\alpha =45.5\pm 0.9\,{\rm mas\,yr}^{-1}, \quad \mu_\delta = -7.8 \pm 0.9 \,{\rm mas\,yr}^{-1}.
\end{equation}
Excluding the FOC data from consideration, we obtained
\begin{equation}
\label{pmnoFOC}
 \mu_\alpha =45.4\pm 0.9\,{\rm mas\,yr}^{-1}, \quad \mu_\delta = -7.7 \pm 0.9 \,{\rm mas\,yr}^{-1}.
\end{equation}
at the middle epoch 
2013.41.
These proper motion estimates agree reasonably well and show that the systematic error of the FOC cannot be 
large. We therefore chose the proper motion estimate using all three epochs (Equation \ref{pm}) as our final result. The corresponding total proper motion and its position angle (counted from north through east) are
\begin{equation}
    \mu=48.3\pm 0.7 \,\, {\rm mas\,\,yr}^{-1}, 
    \quad\quad {\rm P.A.} = 100.4^\circ \pm 0.9^\circ
\end{equation}
The proper motions along the R.A. and declination, together with other reported proper motion values, are demonstrated in Figures~\ref{fig:pm} and ~\ref{fig:overview}.  
Our measurement is more precise than the values derived from the available radio observations and demonstrates the power of combining the HST data with Gaia  astrometry.
The corresponding transverse velocity of the pulsar is $V_\perp = (80\pm 1) d_{350}$\,km\,s$^{-1}$.
If the distance is indeed around 350\,pc, this suggests either a fairly low kick at birth or a potentially high velocity component along the line of sight. Only 6\% of pulsars with ages $<10$\,Myr are expected to have a 3D-velocity of 80\,km\,s$^{-1}$ \citep{Verbunt2017}.

\subsubsection{Photometry}
\label{sec:photometry}
We use apertures measurements to derive the NIR fluxes.
Since the sky background is rather inhomogeneous, with notable flux enhancements towards the neighbouring bright stars and diffraction spikes, as well as with apparent local noise features (e.g., there is one at $\lesssim 0\farcs05$ north of the target aperture in F110W, see Figure~\ref{fig:apertures} in the Appendix), we have to restrict the apertures to a small size $0\farcs{2}$ and use sky background apertures instead of a sky annulus around the source.
In order to account for systematic errors due to the background choice, we 
use several circular background apertures placed around the pulsar. 
As summarized in Appendix~\ref{photmethods}, we tested different choices of background apertures and methods to determine the sky background and uncertainties. The results agree within uncertainties. In the following we use the fluxes that are obtained with 30 (22) background apertures around the source in  F110W (F160W) (Table~\ref{tab:fluxesB} in Appendix~\ref{photmethods}).
We corrected the measured fluxes for the aperture size using the encircled energy fractions $\phi=0.739$ 
and 0.608 for F110W and F160W, respectively\footnote{WFC3 instrument handbook, \url{https://hst-docs.stsci.edu/wfc3ihb/chapter-7-ir-imaging-with-wfc3/7-6-ir-optical-performance}}. 
Our final NIR flux density estimates for PSR\,B1055$-$52 are $f_\nu = 228\pm 40$\,nJy ($261\pm 17$\,nJy) for F110W (F160W), corresponding to $24.7\pm0.2$\,mag ($24.08\pm 0.07$\,mag) in the Vega magnitude system (all uncertainties are $1\sigma$). 

\subsection{Archival \emph{Fermi} data}
\label{fermidata}
Data from Fermi-LAT were used to extend the energy spectrum of PSR B1055--52 up to GeV energies. We downloaded the Fermi-LAT data in the 100 MeV to 300 GeV energy range from about 13.9 years of observations. The data included a 15$^{\circ}$ region of interest (ROI) centered on PSR B1055--52. We reduced the data using the {\tt Fermipy} software package \citep{Wood2017}, which uses the latest version (v2.2.0) of the {\tt Fermitools}. The data were reduced following the standard procedures and using the {\tt P8R3$\_$SOURCE$\_$V3} version of the response function, the ``Source'' event class (i.e., evclass = 128), and both front and back converting events (i.e., evtype = 3). We then performed a binned analysis (using 4 bins per decade in energy) which allows for the use of the energy dispersion correction.

The initial model for the surrounding sources in the ROI were adopted from the 4FGL-DR3 catalog \citep{Abdollahi2022ApJS260}. 
The Galactic diffuse emission and isotropic emission were accounted for using the {\tt gll$\_$iem$\_$v07.fits} and {\tt iso$\_$P8R3$\_$SOURCE$\_$V3$\_$v1} models, respectively. 
We used the simple
PLEC (power law with exponential cutoff) model to fit PSR B1055-52:
\begin{equation}
\frac{dN}{dE} = K_\gamma \left(\frac{E}{E_0}\right)^{\alpha_\gamma -1} \exp\left(-\frac{E}{E_c}\right)
\label{eqn:gamma-ray_model}
\end {equation}
and found the following fitting parameters at fixed scaling energy $E_0=500$ MeV: 
$K_\gamma=(3.116\pm 0.032)\times 10^{-10}$ photons cm$^{-2}$ s$^{-1}$ MeV$^{-1}$,
$\alpha_\gamma = 0.115\pm0.018$, 
$E_c = 1130\pm 20$ MeV, which correspond to the energy flux $F_{\rm 0.1-100\,GeV}=
(2.76\pm0.02)\times 10^{-10}$ erg cm$^{-2}$ s$^{-1}$.
We will use this result for the discussion of the multiwavelength spectrum of B1055 in Section~\ref{XrayGamma}.

\section{Discussion}
\subsection{X-ray flux and X-ray spectrum}
\begin{figure}[t]
    \includegraphics[width=8.9cm]{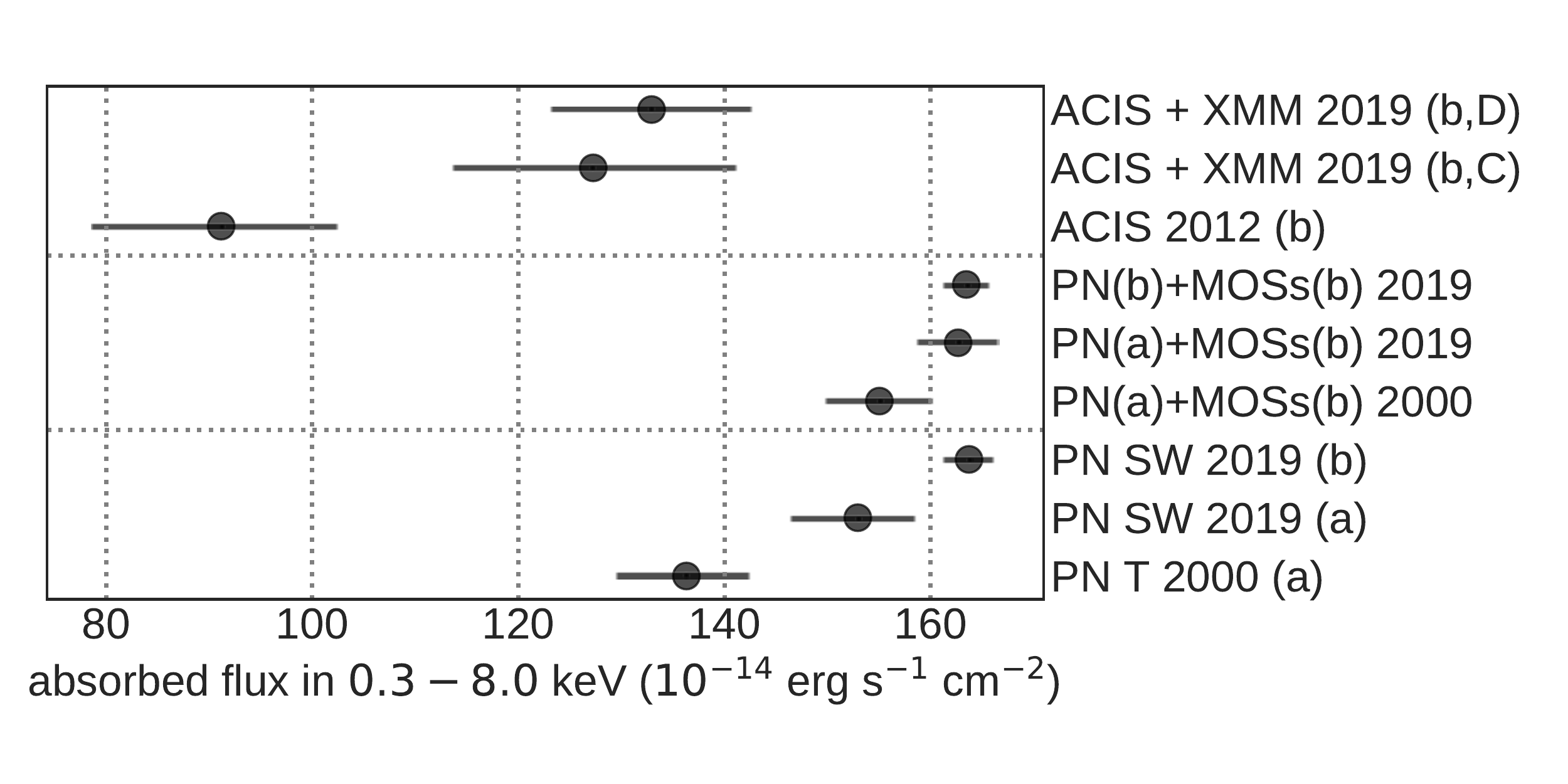}\vspace{-0.2cm}
    \caption{The absorbed fluxes for different data sets as indicated. The energy ranges of the model fits is (a) $0.4-8$\,keV (due to the restricitions of the early PN timing data) and (b) $0.3-8$\,keV. For the tied fits of the ACIS + XMM 2019 data (ACIS + XMM 2010 data give similar results), (C) marks the fit where only $N_{\rm H}$ and the photon index are tied, (D) the fit where all parameters except $N_{\rm H}$ are tied.See text for details.  All shown (statistical) uncertainties are $3\sigma$. }
    \label{fig:XMMChandra}
\end{figure}

We compare the absorbed X-ray fluxes in the energy range $0.3-8$\,keV  at different epochs. They are obtained using the \texttt{cflux} model component in XSPEC.
As detailed in Sections~\ref{XMMcomp} and ~\ref{XMMCXOcomp}, we find an acceptable consistency of the absorbed fluxes for the 2000 and 2019 data from XMM-\emph{Newton} but a significant flux deviation for the 2012 \emph{Chandra} data (see Figure~\ref{fig:XMMChandra}). 
A  \emph{short-term} spectral change in 2012, but consistent values in 2000 and 2019 cannot be entirely ruled out. 
A cross-calibration error is another explanation. Using only the data from the two XMM-\emph{Newton} epochs, we summarize the constraints on the phase-integrated X-ray spectrum in section~\ref{XMMspec}.

\subsubsection{Comparison of the two epochs XMM-\emph{Newton} data}
\label{XMMcomp}
We do not list all individual fit parameters here as our results for the 2000 data are consistent with those reported by \citet{Posselt2015}, and those for the 2019 data are consistent with those reported by \citet{Rigoselli2022}. 
Note that since the 2000 PN timing data only start at 400\,eV, $0.4-8$\,keV  is used to fit these data, and the flux in the $0.3-8$\,keV interval is estimated based on the model using the \texttt{cflux} component in XSPEC. 
This is indicated by (a) in Figure~\ref{fig:XMMChandra}, in contrast to (b) if data in the energy range $0.3-8$\,keV are available.
Considering only PN, the absorbed fluxes are $F_{\rm abs, 0.3-8 keV} = 136^{+6}_{-7} \times 10^{-14}$\,erg\,cm$^{-2}$\,s$^{-1}$ and $F_{\rm abs, 0.3-8 keV} = (164 \pm 4) \times 10^{-14}$\,erg\,cm$^{-2}$\,s$^{-1}$
for the 2000 and 2019 PN data, respectively ($3\sigma$ uncertainties). The obtained model parameter differ in $N_{\rm H}$ ($93^{+28}_{-22}(1\sigma) \times 10^{19}$ cm$^{-2}$ in 2000 and $19 \pm 5 (1\sigma) \times 10^{19}$ cm$^{-2}$ in 2019) whose value is most sensitive to the missing low energies of the timing mode data. Restricting also the 2019 fit to the $0.4-8$\,keV energy range, increases the model $N_{\rm H}$ ($43^{+22}_{-18} \times 10^{19}$ cm$^{-2}$), and the resulting absorbed flux is  $F_{\rm abs, 0.3-8 keV} = 153^{+6}_{-7} \times 10^{-14}$\,erg\,cm$^{-2}$\,s$^{-1}$ ($3\sigma$). It is closer to the 2000 PN value, but has a $5\sigma$ difference.\\

Including the MOS data (all parameters tied for the three EPIC instruments except a multiplicative factor, similarly as described by \citealt{Posselt2015}) leads to agreement of all spectral fit parameters of the 2000 and 2019 EPIC data within $1\sigma$ except for $N_{\rm H}$ where the difference is slightly larger. The absorbed fluxes values agree at the $4\sigma$ or $3\sigma$ level depending on whether the lower energy boundary for the 2019 PN data set is 0.3\,keV or 0.4\,keV (see Figure~\ref{fig:XMMChandra}). 
Considering the restricted energy range in the 2000 PN data, the respective missing offset map, as well as increased background due to its timing mode, we regard the absorbed fluxes of the individual PNs as well as the EPIC (PN+2MOS in each epoch) as marginally consistent values. The flux difference of about $\sim +5$\% between 2000 and 2019 is nowhere near the $\sim -30$\% 
that was seen between the 2000 EPIC data and 2012 ACIS data by \citet{Posselt2015}. From a long-term study  of the isolated neutron star RX\,J1856.5$-$3754, \citet{DeGrandis2022} found small position-dependent effects for the derived spectral parameters for PN data. They assumed an additional systematic error of 1\%. If such a systematic error is included, the absorbed fluxes from 2000 and 2019 are consistent within $3\sigma$. 
Hence, we conclude that the consistency of the X-ray fluxes and spectral parameters for the PN-only and the EPIC data do not support the hypothesis of respective long-term changes of this pulsar. 
It is noteworthy that restricting the lower energy boundary increases the derived $N_{\rm H}$ and decreases the estimates for the absorbed flux as a consequence.\\

\subsubsection{Comparison of the \emph{Chandra} and XMM-\emph{Newton} data}
\label{XMMCXOcomp}
Using an updated calibration database (CALDB 4.9.4\footnote{Subsequent CALDB changes do not affect the 2012 data, but are updates for the more recent \emph{Chandra} data.}), we fit the 2012 \emph{Chandra} ACIS data in the energy range $0.3-8$\,keV. The absorbed flux is $F_{\rm abs, 0.3-8 keV} = 91^{+11}_{-12} \times 10^{-14}$\,erg\,cm$^{-2}$\,s$^{-1}$ ($3\sigma$), a $\sim 10\sigma$  difference considering the lowest XMM-Newton values (PN 2000). 
Tying the ACIS data to the 2019 EPIC data (similar results for the 2000 data) decreases the difference. 
If all model parameters are tied, there are strong systematic residuals of the ACIS data at low energies ($<1$\,keV) pointing towards the region of main difference. We considered therefore two other variants for the tied model fits.
In the first one, $N_{\rm H}$ and the photon index are tied (C in in Figure~\ref{fig:XMMChandra}), 
as in \citet{Posselt2015}, 
while in the second one all model parameters are tied except  $N_{\rm H}$ (D in in Figure~\ref{fig:XMMChandra}). Both fit variants bring the ACIS absorbed flux closer to agreement with the 2000 PN-only fit, but the difference to the 2019 EPIC fit still has a significance of $\gtrsim 8\sigma$. $N_{\rm H}$ from the ACIS fit (D) is a factor 2.5 higher than the EPIC-derived value. 
Similar to the 2000 PN data, where the low energy events were not available, the larger estimated value for  $N_{\rm H}$ results in a lower flux. 
This, and the strong systematics at soft energies if all parameters are tied, could indicate that the ACIS contamination layer and imperfect response correction at low energies may cause a larger than usual cross-calibration uncertainty for this 
soft X-ray source. Since, the ACIS response is known to change over time due to the accumulating contamination, potential calibration observations  for comparisons with XMM-Newton need to be close in time.\\   

We checked the 2011$-$2013 \emph{Chandra} calibration observations of RX\,J1856.5$-$3754 for a potential comparison with the XMM-Newton results by \citet{DeGrandis2022}. However, 
being obtained in a different instrument mode (ACIS-S/LETG or HRC/LETG 
versus the ACIS-I image of B1055), 
they are not suited as a cross-calibration check for 2012. The Geminga pulsar, another of the Three Musketeers, was observed with ACIS-I in 2012$-$2013. Table~2 by \citet{Posselt2017} showed a comparison of the spectral constraints on the pulsar from the ACIS-I observations in comparison to the 2002 XMM-Newton constraints from \citet{deLuca2005}. For fixed $N_{\rm H}$ value, there is a similar tendency of higher blackbody temperatures for the ACIS data as we saw in the B1055 comparison. However, Geminga has a prominent pulsar wind nebula (B1055 and RX\,J1856.5$-$3754 do not) which is time-variable and which is more blurred with the pulsar emission in the XMM-Newton image. For these reasons, Geminga cannot be used as a good quantitative cross-calibration check.

\subsubsection{The phase-integrated X-ray spectrum}
\label{XMMspec}
With the consistency of the X-ray fluxes and spectral parameters established, we consider the entire XMM-\emph{Newton} EPIC data set (twelve spectra), firstly, carrying out a fit for only the PL  spectral component at energies $>3$\,keV as described in Section~\ref{sec:obsxmm}.
The derived photon index $\Gamma = 1.57^{+0.26}_{-0.25}$ is consistent with (but lower than) 
$\Gamma\sim 1.9$ which is obtained if lower energies are included and all parameters are allowed to vary freely. 
The fit results for the twelve spectra with fixed $\Gamma = 1.57$ in Table~\ref{XMMresult} agree within uncertainties with previous XMM-{\sl Newton} results from \citet{Rigoselli2022,Posselt2015}, and \citet{deLuca2005}.
Our final $N_{\rm H}=14^{+3}_{-2} \times 10^{19}$ cm$^{-2}$ is not surprisingly dominated by constraints from the additionally available soft energy events in the 2019 data.

\subsection{UVOIR spectrum}
\begin{figure}
\begin{center}
\includegraphics[width=8.2cm]{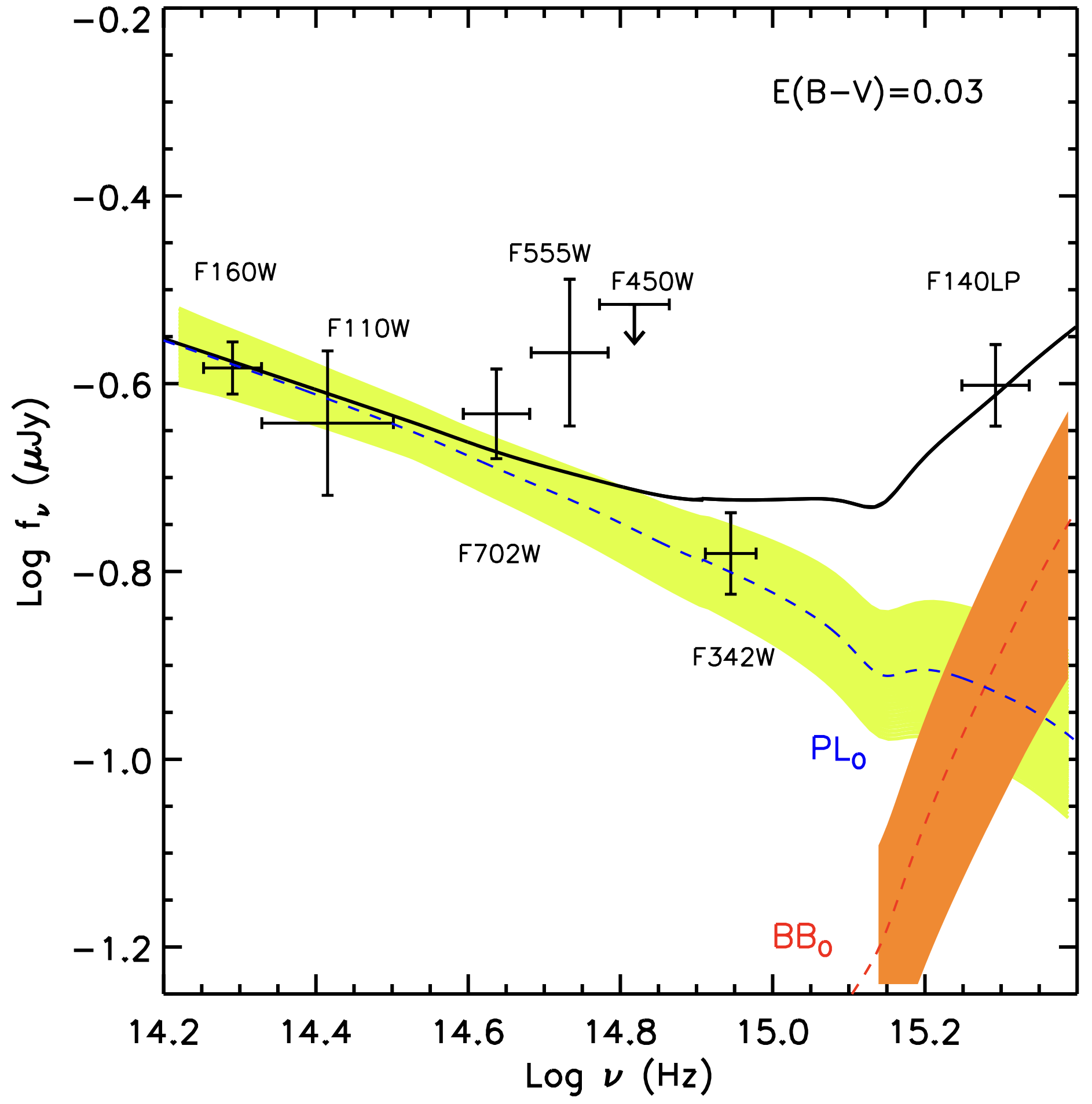}
\end{center}
\vspace{-0.2cm}
\caption{The measured NIR-FUV flux densities (not corrected for reddening) and the (absorbed) PL$_O$+BB$_O$ fit for $E(B-V)=0.03$. The $1\sigma$ uncertainty range for the PL$_O$ component (blue dashed line) is indicated by the green shaded area, the $1\sigma$ uncertainty for the BB$_O$ component (red dashed line) in the UV is shown with the orange shaded area. All measurement uncertainties here and in Figures \ref{fig:alphaTebv} -- \ref{fig:IROUVxG} are $1\sigma$ uncertainties.}
\label{fig:IROUV}
\end{figure}

The new NIR flux densities are plotted together with the previous optical and UV measurements by \citetalias{Mignani2010} and \citet{Mignani1997} in Figure~\ref{fig:IROUV}. The plotted F140LP (FUV) flux density is a factor of 1.24 lower than shown   
by \citetalias{Mignani2010} in their Figures\,2 and 3 because we took into account a recent correction to the ACS/SBC sensitivity \citep{Avila2019}. The two-component PL$_O$+BB$_O$ model 
can be constrained once a proper value for the reddening is taken into account.  

\begin{figure*}
\begin{center}
\includegraphics[height=5.3cm]{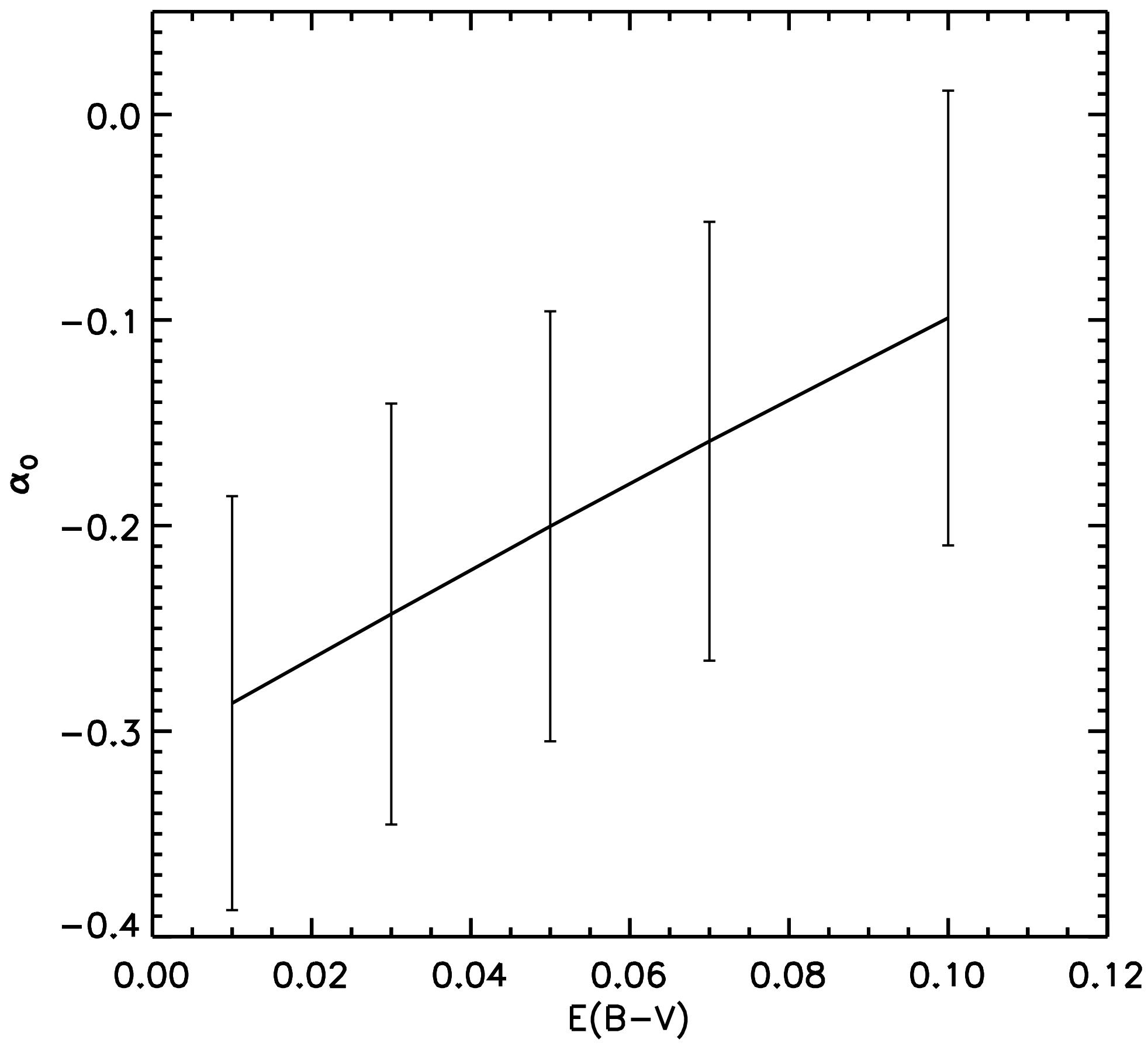}
\includegraphics[height=5.3cm]{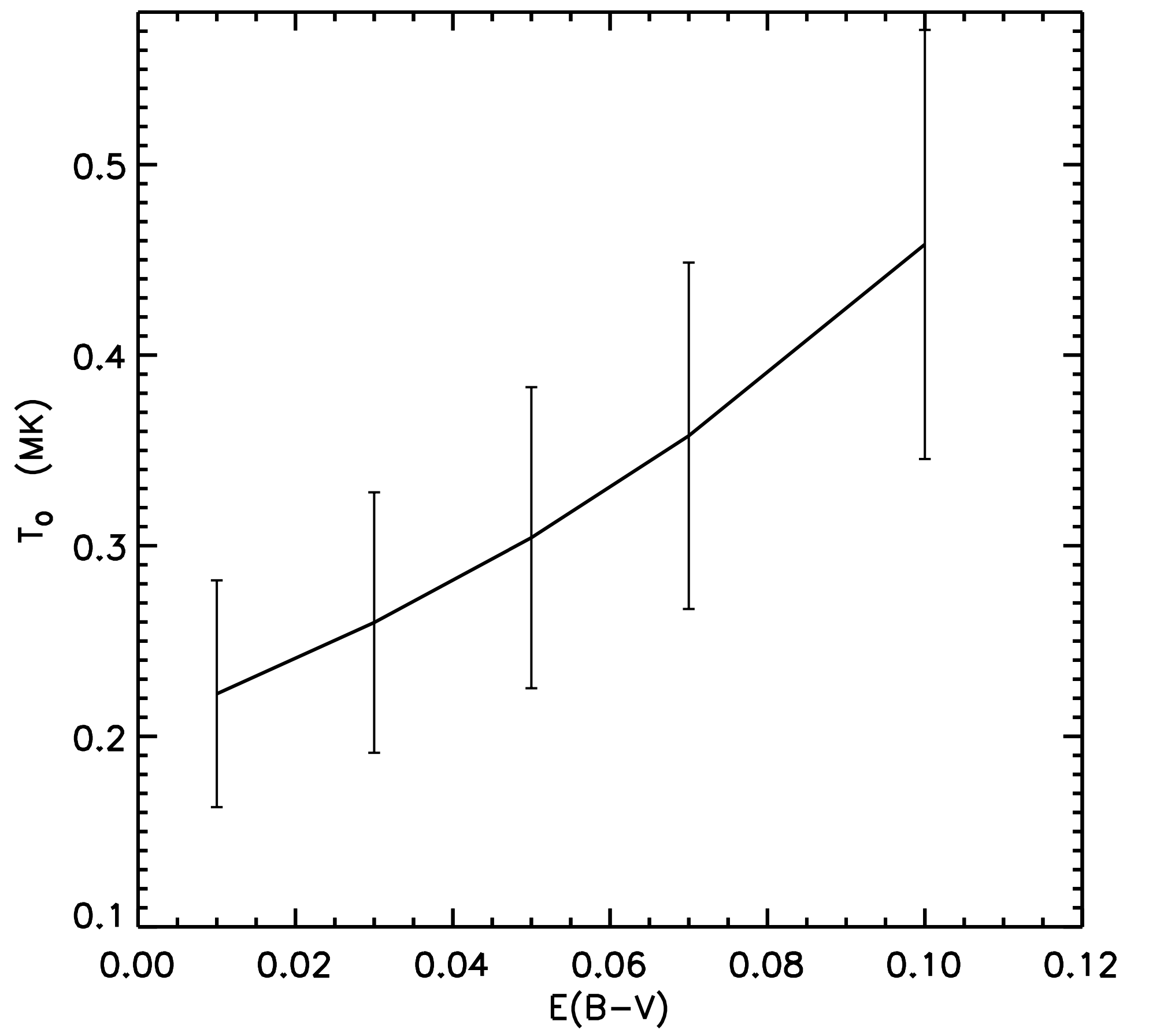}
\includegraphics[height=5.3cm]{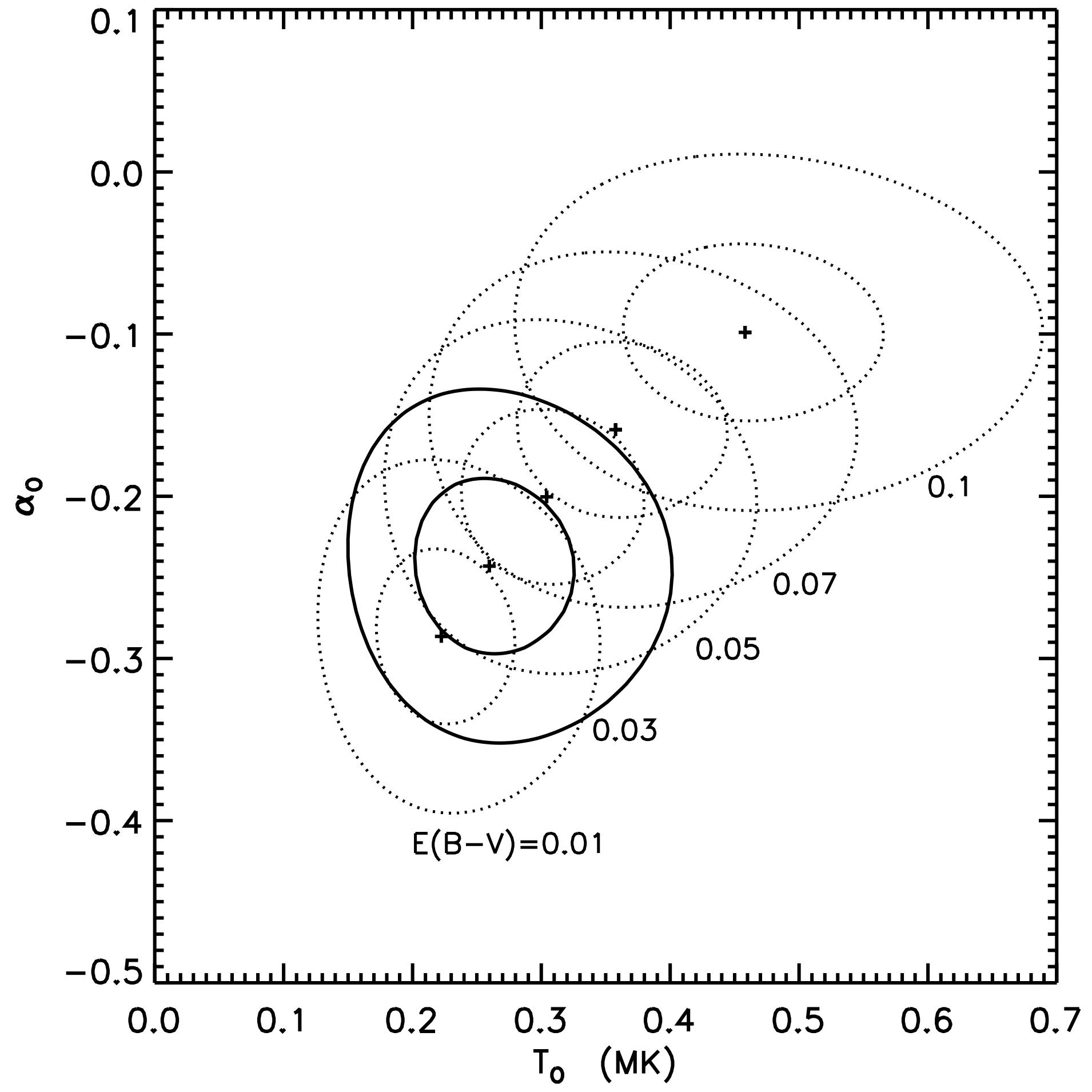}
\end{center}
\vspace{-0.2cm}
\caption{The correlation of the PL$_O$ slope, UV brightness temperature, and $E(B-V)$. The \emph{left}  and \emph{middle panels} show the dependency of $\alpha_O$ and the UV temperature $T_O$ on $E(B-V)$, respectively. The \emph{right panel} shows the correlation of the slope and temperature for different $E(B-V)$. Uncertainties are $1\sigma$, contour levels correspond to 68\% and 99\% confidence levels.}
\label{fig:alphaTebv}
\end{figure*}

\subsubsection{Limits on extinction toward PSR B1055--52}
\label{extinct}
Based on $N_H= 2.7\times 10^{20}$ cm$^{-2}$ \citep{deLuca2005} and the correlation between $N_H$ and visual extinction $A_V$ suggested by \citet{Predehl1995},  \citetalias{Mignani2010} assumed a reddening value of $E(B-V)= A_V/3.1 = 0.07$\footnote{$E(B-V)$ is in mag as usual. We omit mag for easier reading.}. 
As the hydrogen column density from the new XMM-Newton data is a factor of 2 lower, $N_H=1.4^{+0.3}_{-0.2}\times 10^{20}$ cm$^{-2}$, we estimate 
$E(B-V) \approx 0.025$, 0.020, and 0.016 from the empirical relations by \citet{Predehl1995, Gorenstein1975} and \citet{Foight2016}, respectively. 
If the dispersion measure of 
B1055, ${\rm DM} = 29.69$\,cm$^{-3}$\,pc$^{-1}$ \citep{Petroff2013}, is used together with the empirical relation between DM and $N_H$ \citep{He2013}, one would expect  a larger $N_H=8.9^{+3.9}_{-2.7}\times 10^{20}$ cm$^{-2}$ 
that would in turn lead to larger reddening estimates, e.g., $E(B-V)=0.129$  
with the relation from \citet{Gorenstein1975}. However, the DM--$N_H$ relation is rather uncertain, and the free-electron-to-dust ratio can be quite different for different paths in the ISM.
We regard the XMM-\emph{Newton} derived $N_H$ as the better constraint on the reddening. However, all the empirical relations have quite some scatter, the $N_H$ itself depends on the choice of the abundance tables and spectral model, and $R_V$ depends on the line of sight. Therefore, the obtained extinction value is not certain. 
Because the distance of PSR\,B1055--52 is not well known, the constraints from 3D extinction models are limited. The recent model by \citet{Lallement2022} indicates a nominal $E(B-V)=0.03$
at a distance of 270\,pc, 0.047
at 350\,pc, and 0.1\,
at 780\,pc (R. Lallement, personal communication\footnote{The extinction cube is accessible via \href{https://explore-platform.eu/}{https://explore-platform.eu/} }), however uncertainties in $E(B-V)$ can easily reach $\pm 0.02$
in this direction of sight, according  to the online tool by \citet{Capitanio2017}. 
Based on the new XMM-\emph{Newton} data, we choose $E(B-V)=0.03$ for our plots in the following, 
but we investigated the range $E(B-V)=0.01$--0.1 in our fits of the UVOIR broad-band spectra.

\subsubsection{Fits of UVOIR broad-band spectra}
We fit the 
flux density values measured in six filters with the PL$_O$+BB$_O$ model\footnote{MPK10 used the 
RJ approximation for the thermal component. Since our best-fit temperatures are lower, we use a more general Planck spectrum.}:
\begin{equation}
f_\nu^{\rm mod} = \left[f_{0} \left(\frac{\nu}{\nu_0}\right)^{\alpha_O} + \frac{R_O^2}{d^2} \pi B_\nu(T_O)\right] \times 10^{-0.4 A_\nu}\,,
\label{eq:pl+rj_absorbed}
\end{equation}
where 
$\nu_0$ is the reference frequency for the PL component (we choose $\nu_0=1\times 10^{15}$ Hz, which corresponds to $\lambda_0=3000$ \AA), 
$d= 350\,d_{350}\,{\rm pc}$ is the distance, $R_O = 15\,R_{O,15}\,{\rm km}$ is the NS radius\footnote{We use a reference NS radius of 15\,km (instead of the 13\,km in \citetalias{Mignani2010}) because recent estimates (e.g., \citealt{Raaijmakers2021,Salmi2022})
indicate larger NS radii as more likely.}, 
$T_O = 10^6 T_{O,6}\,{\rm K}$ is the NS surface temperature (both the radius and the temperature are as measured by a distant observer), $B_\nu(T_O)$ is the Planck function, and $A_\nu$ is the extinction coefficient, proportional to the color excess (reddening) $E(B-V)$ \citep{Cardelli1989}.
We consider $f_{0}$, $\alpha_O$ and $T_O$ as fitting parameters, and use the chi-square minimization to find the best-fit parameter values and their uncertainties at fixed values of the color excess 
and radius-to-distance ratio (see \citealt{Abramkin2022} for details).
In our fits we assume $R_{O,15}/d_{350} = 1$ unless stated otherwise. 
An example of the PL$_O$+BB$_O$ fit (at $E(B-V)=0.03$) is shown in Figure \ref{fig:IROUV}. The minimum chi-square value is 
$\chi^2_{\rm min} = 5.6$ for three degrees of freedom.
The fit is marginally acceptable considering the involved measurement uncertainties, with the largest residual in the optical F555W band. Note that this fit is significantly better than a PL-only fit, for which
$\chi^2_{\rm min} = 14.1$ for four degrees of freedom.
The best-fit values and uncertainties of the slope $\alpha_O$ and temperature $T_O$ for different values of $E(B-V)$ are shown in Figure \ref{fig:alphaTebv}.
They both grow with increasing $E(B-V)$, and show a weak anti-correlation with each other, in particular for low $E(B-V)$.\\

Thanks to the new NIR data, our uncertainty of the PL$_O$ slope is a factor 3 smaller than achieved by \citetalias{Mignani2010}.
At $E(B-V)=0.07$ (the \citetalias{Mignani2010}'s value), we obtain $\alpha_O=-0.16\pm 0.10$, 
which is substantially harder than the old slope of \citetalias{Mignani2010}, $\alpha_O = -1.05 \pm 0.3$.
For $E(B-V)=0.03$ (corresponding best to our $N_{H}$ estimate, Section~\ref{extinct}), we obtain $f_0 = 178\pm 25$ nJy and $\alpha_O=-0.24\pm0.10$. 
The measured PL$_O$ slope is comparable to those of the majority of pulsars detected in the IR-optical, 
$-0.5\lesssim \alpha_O\lesssim +0.2$ (e.g., \citealt{Mignani2007}),
including the middle-aged pulsars
B0656 ($\alpha_O=-0.41\pm0.08$), and Geminga ($\alpha_O= -0.46\pm 0.12$) \citep{Kargaltsev2007}. 
The NIR-UV flux in the (de-reddened) PL$_O$ component, 
$F^{\rm PL}_{\rm unabs}(0.2$--$2\,\mu{\rm m}) = 2.6 \times 10^{-15}$ erg cm$^{-2}$ s$^{-1}$, 
corresponds to the nonthermal luminosity
$L^{\rm PL}(0.2$--$2\,\mu{\rm m})= 
4\pi d^2 F^{\rm PL}_{\rm unabs}(0.2$--$2\,\mu{\rm m}) =  3.9 \times 10^{28} d_{350}^2$ erg s$^{-1}$,
 assuming isotropic emission. 
 At $d= 350$ pc, the NIR-UV efficiency is 
$\eta_O= L^{\rm PL}(0.2$--$2\,\mu{\rm m})/\dot{E} = 1.3\times 10^{-6}$.
This efficiency is comparable to those of other pulsars from which nonthermal optical emission was detected (e.g., \citealt{Zavlin2004}; \citealt{Zharikov2006}).\\

As one may expect from the anti-correlation between PL slope and temperature in Figure~\ref{fig:alphaTebv}, our larger $\alpha_O$ implies a lower brightness temperature, $T_O=0.36\pm 0.09$\,MK, 
at \citetalias{Mignani2010}'s $E(B-V)=0.07$, in comparison to their 
$T_{\rm RJ}=0.49\pm0.08$\,MK,
at $d_{350}/R_{15}=1$). 
For our choice of $E(B-V)=0.03$, 
we obtain
$T_O = (0.26\pm0.07)$\,MK.
For the adopted distance to B1055, $d=350$ pc, its estimated brightness temperature is slightly lower than those for the presumably younger pulsars B0656 and Geminga that have $T_{\rm RJ} = (0.38\pm 0.04) (d_{293}/R_{15})^2$\,MK and $(0.36\pm 0.02)(d_{250}/R_{15})^2$\,MK, respectively (see \citealt{Kargaltsev2007}) estimated in a similar way. However, 
the brightness temperature of B1055 would be higher if $d>350$ pc and/or $E(B-V) > 0.03$ (see Figure~\ref{fig:alphaTebv}).\\

\subsubsection{Comparison with the X-ray and $\gamma$-ray spectra}
\label{XrayGamma}
\begin{figure*}
\begin{center}
\includegraphics[height=8cm]{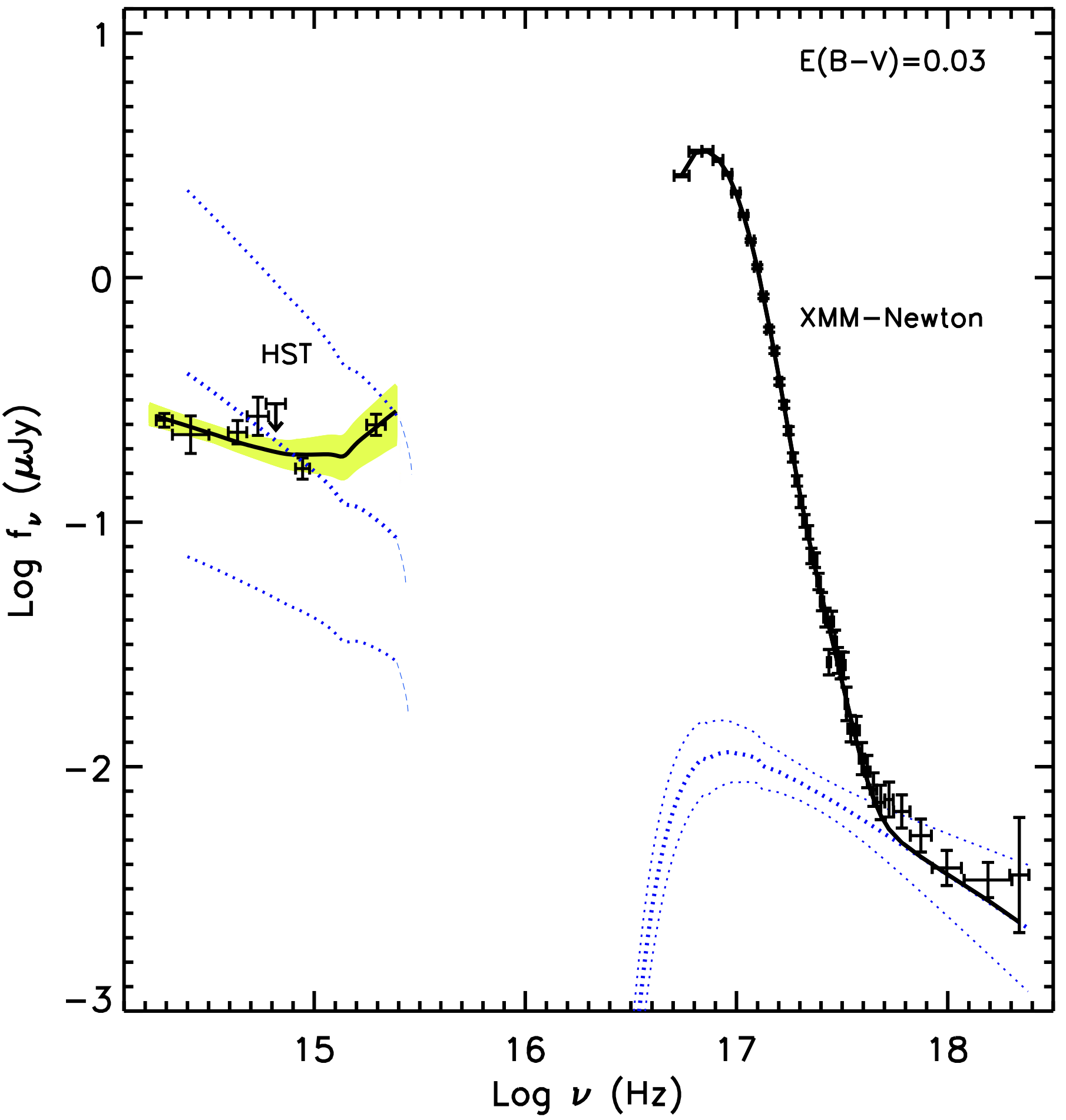}
\includegraphics[height=8cm]{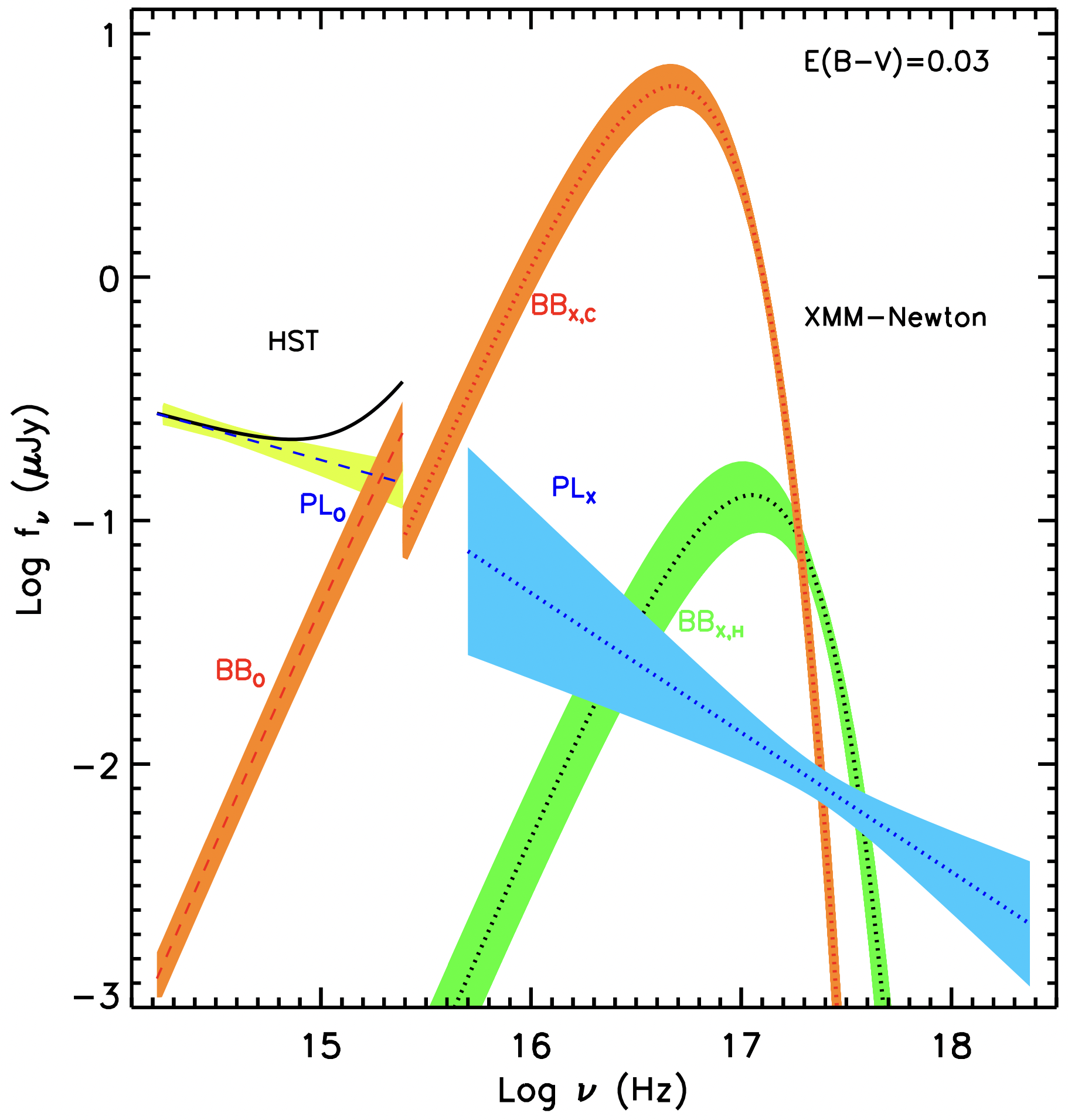}
\end{center}
\vspace{-0.2cm}
\caption{The broad-band spectrum of B1055 from NIR to X-rays. The \emph{left panel} shows the (absorbed) measured flux densities. The \emph{right panel} shows the unabsorbed model components obtained from our fits (see also text). The blue  dotted lines (left panel) or light green (right panel; NIR-FUV) and blue (right panel; X-rays) shaded areas show the PL spectral slopes and their $1\sigma$ uncertainty ranges. The dashed red line and its orange region indicate the tail of the BB$_{\rm O}$ and its $1\sigma$ uncertainty range, obtained from the NIR-UV fit. Similarly, the orange and green areas show the $1\sigma$ uncertainty ranges for the cold and hot blackbody from the X-ray fit  considering the full allowed $1\sigma$ range of the PL. }
\label{fig:IROUVx}
\end{figure*}

\begin{figure*}
\begin{center}
\includegraphics[width=16.8cm]{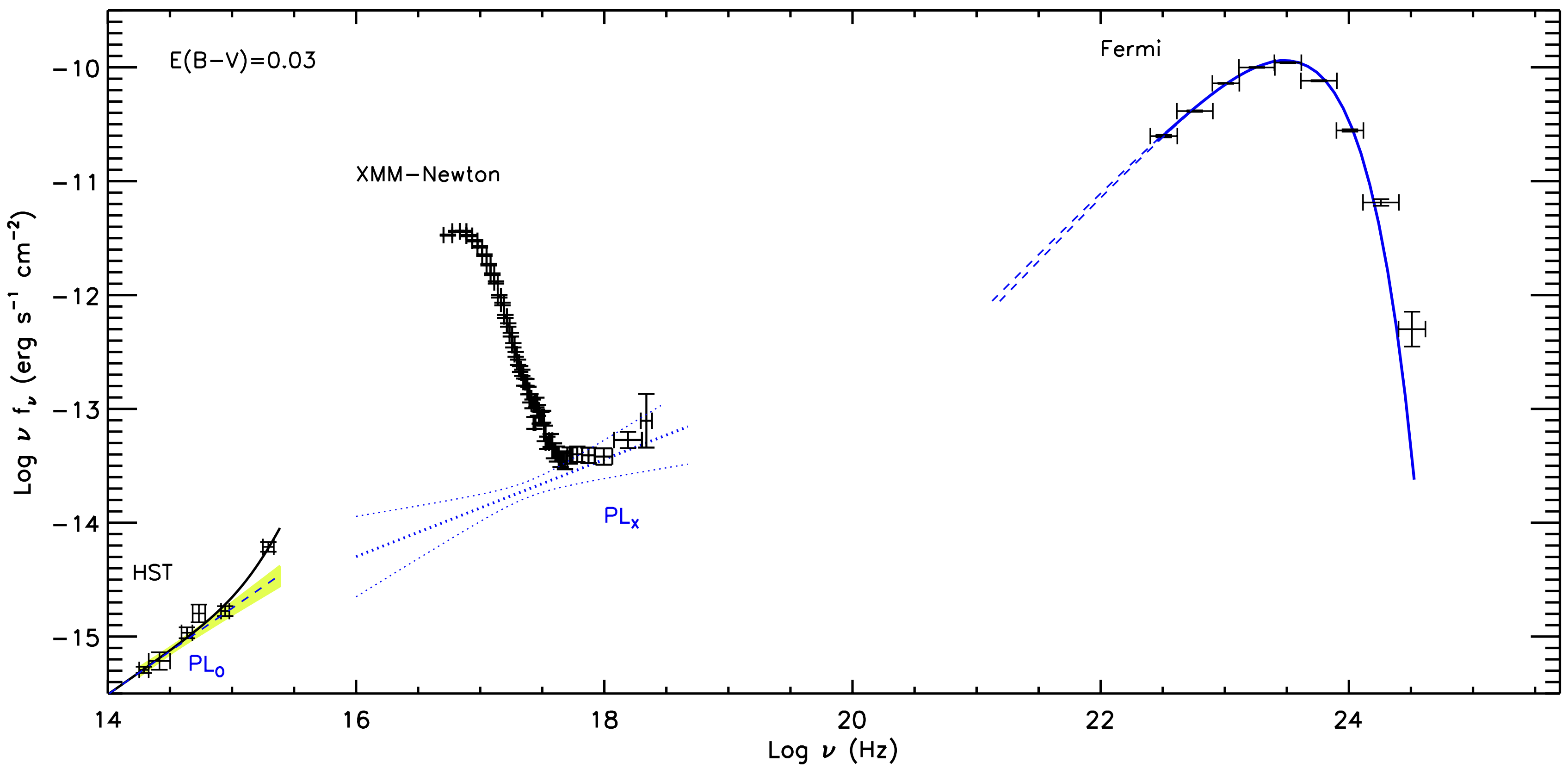}
\end{center}
\vspace{-0.2cm}
\caption{The broad-band spectral energy distribution  of B1055 from NIR to $\gamma$-rays. The NIR-UV measurements and curves correspond to the left panel of Figure~\ref{fig:IROUVx} but for unabsorbed flux densities multiplied with the frequency.
Model constraints on the $\gamma$-ray spectrum and X-ray PL spectral component are indicated with dashed and dotted blue lines, respectively. } 
\label{fig:IROUVxG}
\end{figure*}

The substantial revision of the model parameters changes the relationship between the UVOIR and X-ray spectra.
The PL$_O$ component, for instance, 
now connects relatively smoothly with the X-ray PL$_X$ component, see right panel of Figure~\ref{fig:IROUVx}.
This means that there is no need to invoke a separate source of the nonthermal optical emission, contrary to the hypothesis by \citetalias{Mignani2010}.
The slopes of the 
PL$_O$ and PL$_X$, $\alpha_O= - 0.24\pm 0.10$
and $\alpha_X=-0.6\pm 0.3$, are close to each other. 
However, we cannot exclude that, similar to, e.g., the Geminga pulsar,
the nonthermal component softens in the optical-to-X-ray transition \citep{Kargaltsev2007}. More observations are needed to check this hypothesis.
Similar to other pulsars detected in both X-rays and UV-optical, the nonthermal UVO flux of B1055 is much lower than the nonthermal X-ray one.
The optical-to-X-ray flux ratio, $F^{\rm PL}_{\rm unabs}(0.4$--$0.9\,\mu{\rm m})/F^{\rm PL}_{\rm unabs}(1$--$10\,{\rm keV}) = 0.014$, is close to the 0.011 for B0656 but higher than the 0.002 for Geminga \citep{Zavlin2004}.
Including the NIR measurement for B1055 results in a similar ratio, $F^{\rm PL}_{\rm unabs}(0.2$--$2\,\mu{\rm m})/F^{\rm PL}_{\rm unabs}(1$--$10\,{\rm keV}) = 0.034$.\\

Furthermore, although there still is a noticeable excess (e.g., by a factor of about 2.8 at $\nu=2.5\times 10^{15}$\,Hz)
of the optical-UV thermal component over the extrapolation of the `cold BB' X-ray component, it is 
smaller than the factor of 4.3 found by \citetalias{Mignani2010} in the RJ limit. 
A non-uniform temperature distribution, such that the optical-UV thermal component is emitted from a larger but colder part of the NS surface than the ``cold'' X-ray component, can explain the excess, as already mentioned by \citetalias{Mignani2010}.
The model assumption about two different temperatures, $T_O$ and $T_{X,C}$, is an obvious simplification.
Most likely, there is a continuous temperature distribution over the NS surface, caused by the anisotropic heat conduction in the magnetized NS crust (e.g., \citealt{Geppert2004}).
The  temperature distribution can, in principle, be qualitatively understood from the analysis of pulsations in the X-ray and UV ranges. B1055 has not been investigated for UV-pulsations. In X-rays, a phase-resolved spectral analysis will be presented in a separate work (Vahdat et al., in preparation).  
A caveat for all the temperature and surface luminosity estimates is the (currently very uncertain) distance to the pulsar. This distance should be inferred from parallax measurements.\\

Figure~\ref{fig:IROUVxG} shows the NIR to $\gamma$-ray spectral energy distribution (SED) $\nu f_\nu$, corrected for the interstellar absorption at the optical through X-ray photon energies.
The $\gamma$-ray part of the SED was obtained from our fit of the Fermi-LAT data with the PLEC model 
(see Section~\ref{fermidata}).
In addition to the above-discussed thermal components, which dominate in the 3 eV -- 2 keV band, we see a nonthermal spectrum that covers 10 orders of magnitude in energy, from 1 eV through 10 GeV.  The SED reaches its maximum at $E=(\alpha_\gamma +1) E_c=1.26$ GeV, which indicates that the main energy loss occurs in $\gamma$-rays. Equation~(\ref{eqn:gamma-ray_model}) implies the $\gamma$-ray ``luminosity''
 $L_{\rm 0.1-100\,GeV} = 4\pi d^2 F_{\rm 0.1-100\,GeV}= 4.0\times 10^{33} d_{350}^2$ erg\,s$^{-1}$,
which corresponds to the $\gamma$-ray efficiency $\eta_\gamma = 0.13$\,$d_{350}^2$, much higher than those in X-rays and optical.\\ 
Although the new XMM-\emph{Newton} data allowed us to use only high ($3-10$\,keV) energies for a less biased determination of the PL slope in X-rays, the  slope $\alpha_X = -0.6 \pm 0.3$ (for energies $>3$\,keV) is not much different from previous estimates ($\alpha_X = -0.7 \pm 0.1$ from the 2BB+PL fit by \citet{deLuca2005} for the full XMM energy range).  
The X-ray (and NIR-optical) PL slopes are inconsistent with 
the $\gamma$-ray slope, $\alpha_\gamma = +0.12\pm 0.02$
This means that, most likely, different mechanisms are responsible for the NIR-X-ray and $\gamma$-ray emissions (e.g., synchrotron and curvature radiations\footnote{These two mechanisms can be considered as limiting cases of synchro-curvature radiation (e.g., \citealt{Kelner2015}).}, respectively) or they are generated by different electron-positron populations, perhaps in different parts of the magnetosphere. 
For example, \citet{Cerutti2016} suggested that the most energetic synchrotron $\gamma$-ray emission is produced by particles within the equatorial current sheet beyond the light-cylinder radius.
For the non-thermal X-ray emission, models assume synchrotron emission sites at high altitudes,
 either above the polar cap or in the outer gap (e.g., \citealt{Harding2013}).
Hard X-ray and soft $\gamma$-ray observations, as well as a phase-resolved spectral analysis are the necessary next steps 
to better understand the multiwavelength SED
of this pulsar.

\section{Conclusions} 
Our new XMM-\emph{Newton} and HST observations of PSR\,B1055$-$52
shed new light on its puzzling previous measurement results,
one being the (absorbed) X-ray flux change seen between the 2000 XMM-\emph{Newton} and 2012 \emph{Chandra} observations.
The long-term X-ray properties seem to be stable based on the comparison between the 2000 and 2019 XMM-\emph{Newton} observations. The possibility that short-term X-ray flux changes occurred before or around 2012 cannot be entirely excluded.
A (cross-) calibration issue with the 2012 \emph{Chandra} observation can be another explanation for the 
discrepancies between the latter and the two XMM-\emph{Newton} observations.\\

The NIR data enabled  a new and much tighter constraint on the UVOIR non-thermal PL component,
changing its slope substantially.
The PL components in  the UVOIR and X-ray spectra have similar slopes and connect with each other smoothly, 
suggesting common acceleration and emission mechanisms. 
The slopes of the non-thermal $\gamma$-ray and X-ray spectra are significantly different, which could be due to different emission sites or emission mechanisms. 
NuSTAR observations would help to tighten the constraints on the X-ray PL and elucidate its connection with the $\gamma$-ray emission.\\

The harder PL component of the UVOIR spectrum also implies a lower brightness temperature than previously found by \citetalias{Mignani2010}. Considered together with thermal X-ray spectrum, these different thermal components clearly indicate a non-uniformity of the NS bulk surface temperature.
We note that the B1055's X-ray spectrum is inconsistent with NS atmosphere model spectra, similar to other middle-aged pulsars. This suggests a condensed NS surface, whose spectrum is possibly closer to the BB spectrum.\\ 

The new NIR epoch also allowed us to improve the accuracy 
of the proper motion measurement, with the uncertainty reduced by a factor of 5, and a significant value for its declination component for the first time.
The new astrometric reference frame provided by \emph{Gaia} DR3 proved to be a crucial ingredient for this 
measurement.
In principle, the now well-known proper motion could be used to constrain a kinematic age, providing an independent age estimate preferable to the rather uncertain characteristic age. 
Such information 
enables more reliable comparison with other NSs as well as with theoretical predictions, e.g., NS cooling curves. The large uncertainty of the distance, however, hinders this endeavor as well as firm estimates of luminosities and efficiencies.   
A parallactic distance would be very valuable for a pulsar with such rich multiwavelength phenomenology.\\ 

\begin{acknowledgements}
We thank R. Lallement for supplying us with the reddening information.
We also thank M. Lower, F. Jankowski, and S. Johnston for answering our questions regarding the radio timing solutions.
We thank the referee whose comments helped us to improve the quality of the paper.\\
Support for this work was provided by the National Aeronautics and Space Administration through the XMM-Newton award 80NSSC20K0806. Support for the Hubble Space Telescope program \#15676 was provided by NASA through a grant from the Space Telescope Science Institute, which is operated by the Association of Universities for Research in Astronomy, Inc., under NASA contract NAS 5-26555. 
JH acknowledges support from an appointment to the NASA Postdoctoral Program at the Goddard Space Flight Center, administered by the ORAU through a contract with NASA.\\
This research has made use of the Mikulski Archive for Space Telescopes, USA, and of the VizieR catalogue access tool, CDS, Strasbourg, France.

\facilities{XMM-Newton(EPIC), HST(WFC3/IR)}

\software{XMM SAS \citep{Gabriel2004}, CIAO \citep{Fruscione2006},  Fermipy software package \citep{Wood2017}, XSPEC \citep{Arnaud1996},
DrizzlePac (\url{http://drizzlepac.stsci.edu/}), Graphical Astronomy and Image Analysis Tool \citep{Currie2014},
Python packages: \texttt{ASTROPY}  \citep{astropy}, 
\texttt{MATPLOTLIB} \citep{matplotlib}, 
\texttt{PANDAS} \citep{pandas,reback2020pandas}}

\end{acknowledgements}

\bibliography{B1055bib}{}
\bibliographystyle{aasjournal}

\appendix
\section{Minimizing model correlations for the X-ray photon index}
\label{APLfit}
\begin{figure*}[t]
\begin{center}
\includegraphics[height=7cm]{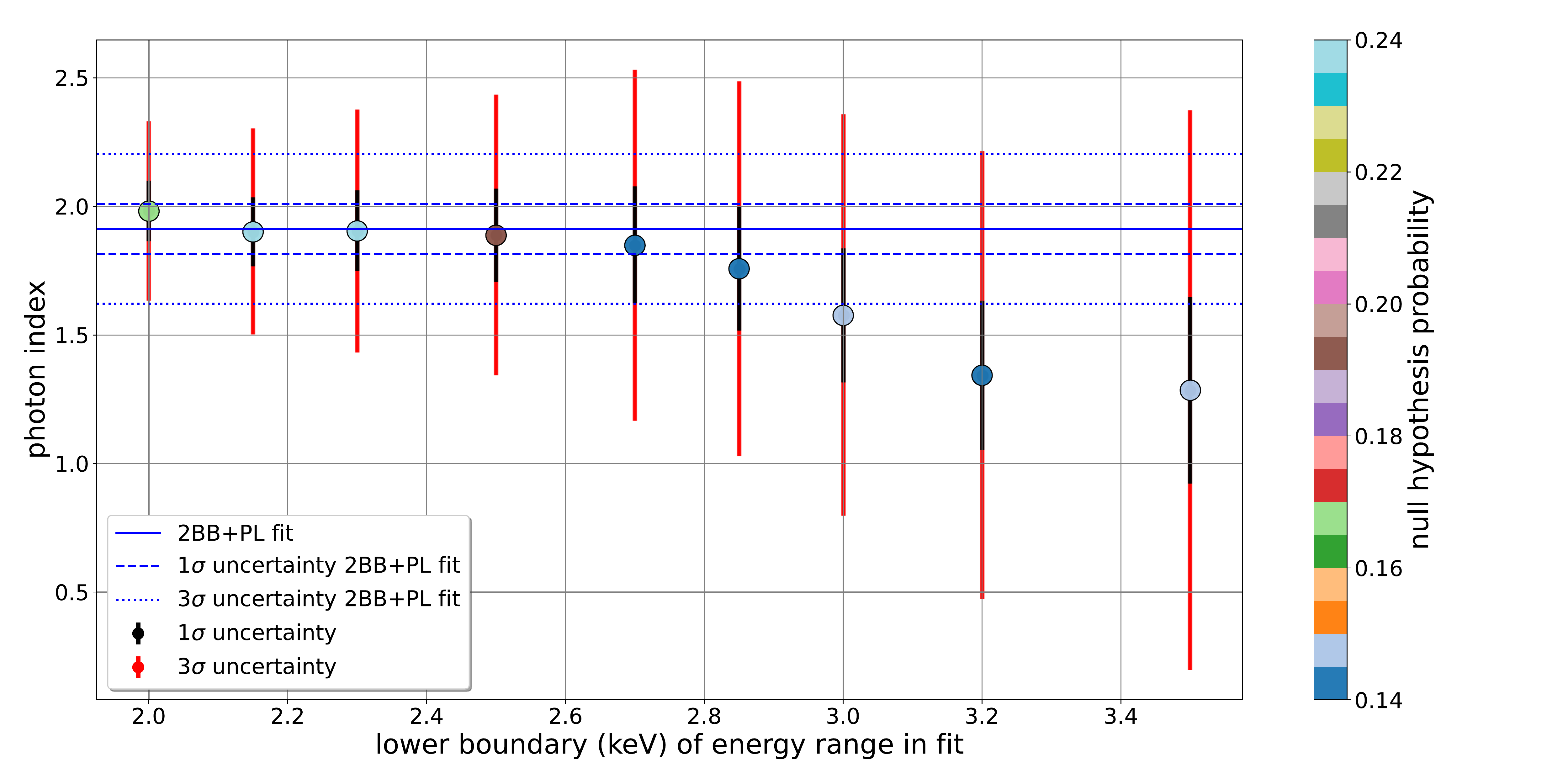}
\end{center}
\caption{The influence of the applied lower energy boundary on the photon index in the PL-only fit. The upper boundary is 10\,keV. For comparison, the photon index result (and its uncertainties) from the 2BB+PL fit in the $0.3-10$\,keV range is indicated with blue lines.\label{PLonlylowE}}
\end{figure*}

\begin{figure*}[]
\begin{center}
\includegraphics[height=9cm]{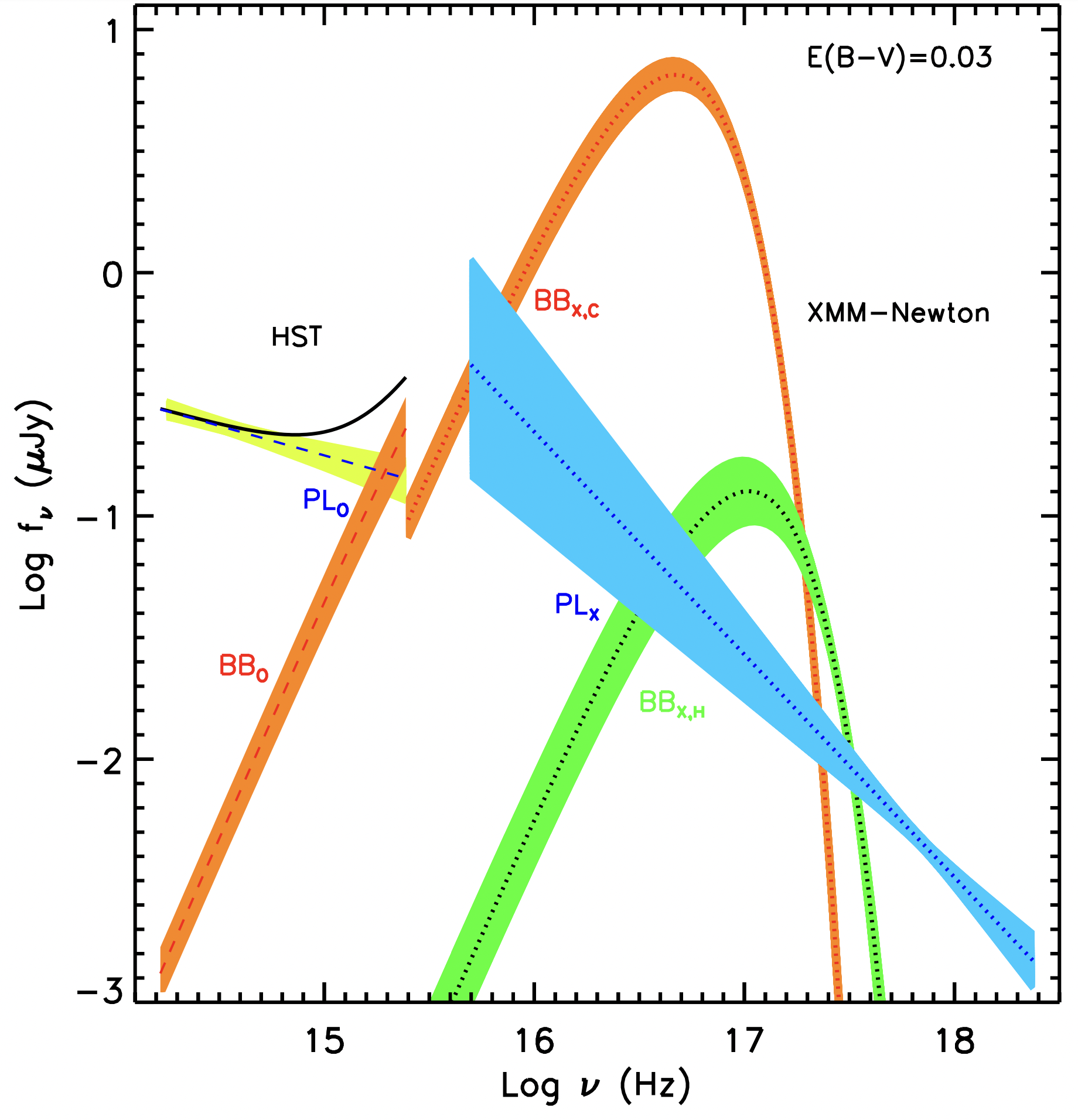}
\end{center}
\caption{The unabsorbed model components obtained from our fits for comparison with Fig.~\ref{fig:IROUVx} (the same colors and lines are used for the same components, see Fig.~\ref{fig:IROUVx} for detailed description). In contrast to the two-step X-ray fit method used for Fig.~\ref{fig:IROUVx}, here, all parameters for the 2BB+PL X-ray model were fit at once.\label{fig:IROUVxPLfree}}
\end{figure*}

As described in section~\ref{sec:obsxmm}, we derive the photon index in our 2BB+PL X-ray spectral model from a PL-only fit using the X-ray energies above 3\,keV. This assumes that the respective photons come from a single non-thermal spectral component, and it aims to minimize effects due to the model-inherent correlations with the parameters of the thermal components. As photons are sparse at higher energies, we chose the lower energy boundary, 3\,keV, based on a compromise between sufficient count statistics and noticable deviation from the full 2BB+PL fit in the $0.3-10$\,keV range, see Figure~\ref{PLonlylowE}.       
Due to the relatively poor count statistics at high X-ray energies, all the resulting fit parameters are still statistically consistent with each other, however with slightly shifted best-fit values. This is also reflected by the fit statistics:  reduced $\chi^2=  1.16$ for 825 dof. for the 2BB+PL fit with fixed photon index, while $\chi^2=  1.15$ for 824 dof, i.e., practically the same, for the full 2BB+PL fit with free photon index.  
For the broadband SED, we also evaluated the influence of the fit results from the one-step 2BB+PL fit with free photon index, see Figure~\ref{fig:IROUVxPLfree}. However, we regard the  results based on the two-step fit method (Fig.~\ref{fig:IROUVx}) as less influenced by parameter correlations.   

\section{Aperture flux density measurement methods}
\label{photmethods}
Since the background is very non-uniform  in the vicinity of the source, we used, as a consistency check of our measurements, two different methods to estimate the background and noise. 
\begin{figure*}[t]
\includegraphics[height=3.75cm]{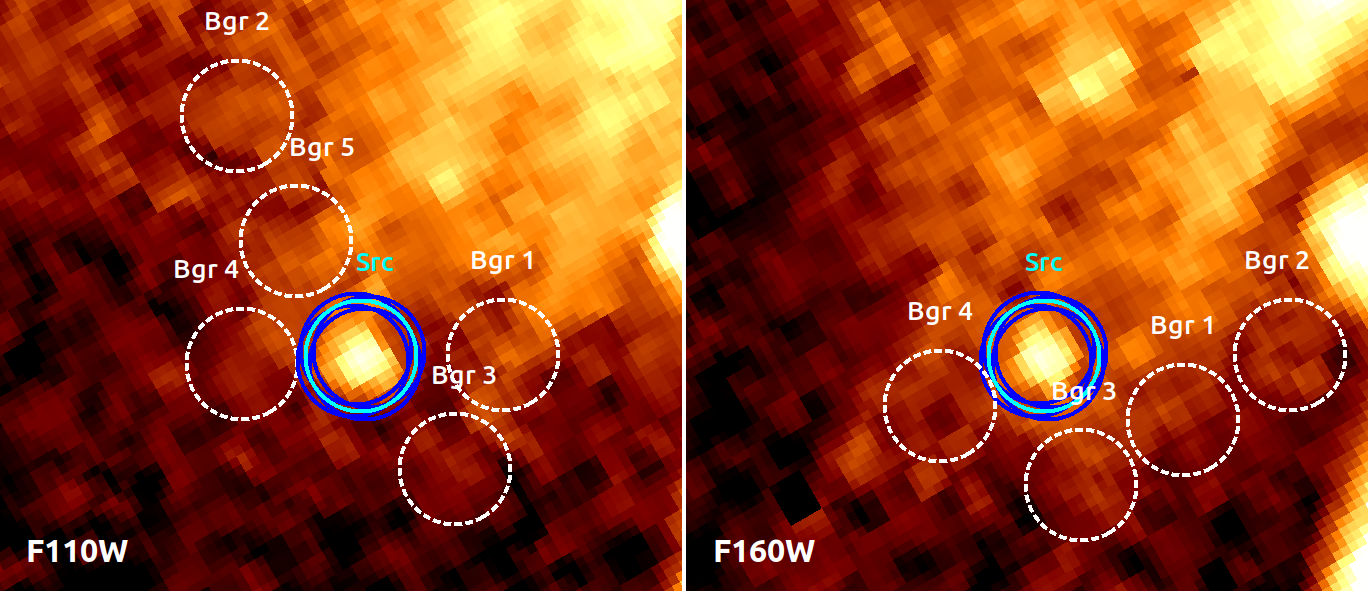}
\hspace{0.3cm}
\includegraphics[height=3.75cm]{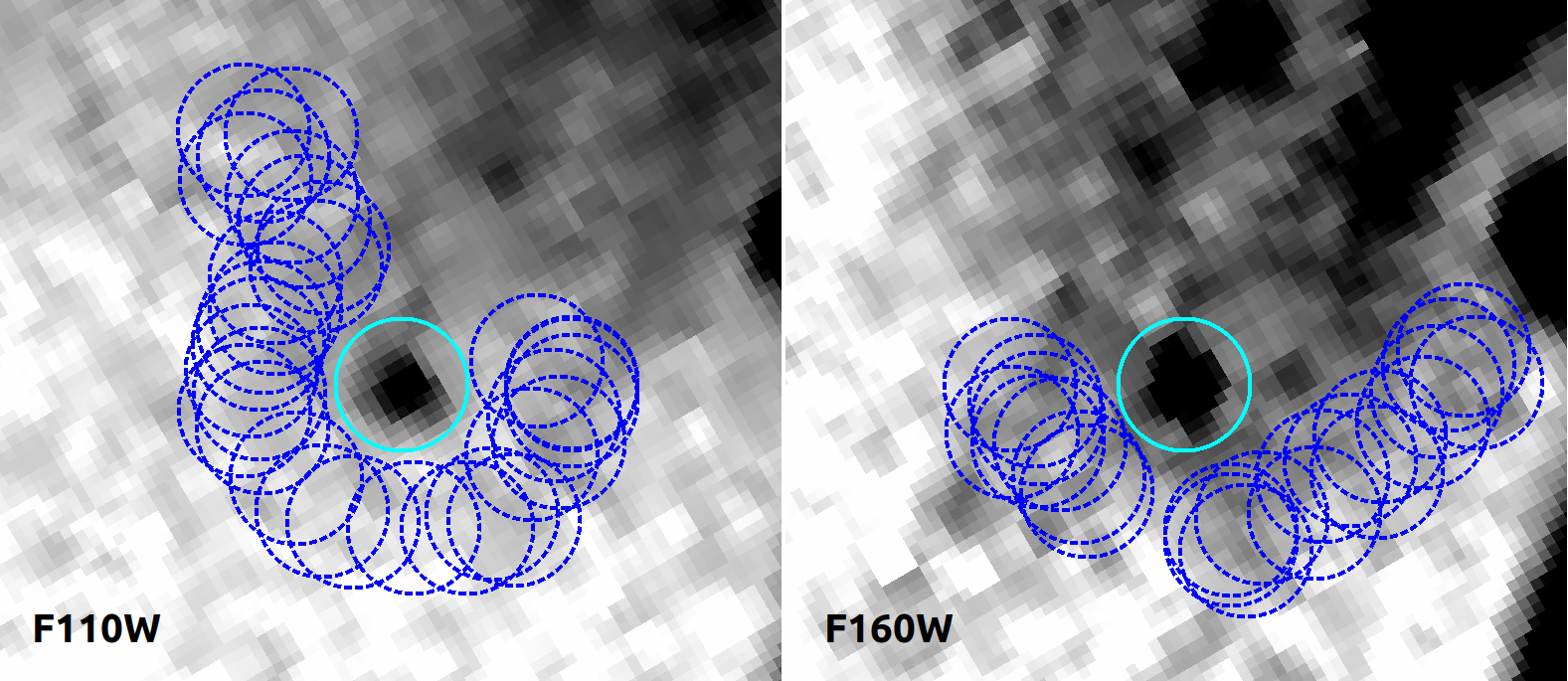}
\caption{The source (solid lines) and background apertures (dashed lines) considered for F110W and F160W photometry of PSR\,B1055$-$52. Each aperture has a radius of $0\farcs{2}$.
The two left panels show the 5 source apertures (cyan and dark blue) and non-overlapping background apertures (white) that were used in our first flux measurement method. The two right panels show the 30 (F110W) and 22 (F160W) background apertures (blue) with one target aperture that were used in our second flux measurement method. See decription in Appendix~\ref{photmethods}.
\label{fig:apertures}}
\end{figure*}
In our first method, we 
use five (four) non-overlapping circular background apertures 
in the F110W (F160W) filters, shown in two left panels of Figure \ref{fig:apertures}.
The source aperture is centered on the respective centroid position (Section~\ref{sec:astrometry}).
To check and account for systematic errors due to centroiding uncertainty, 
we also shift the source aperture by $\pm 1$ pixel ($\pm 0\farcs03$) in either x or y
direction of the image, 
encompassing $5\sigma$ of the radial centroiding uncertainty (Table~\ref{positions}).
For each background aperture, we 
measure the variance, $\sigma^2_b$,
of counts per pixel.
For each combination of source and sky apertures, $i$, we determine the background-subtracted source counts, $N_{s,i}$, and estimate the errors with the following formula: 
\begin{equation}
\label{photm1}
\sigma_{N_{s,i}} = \left[N_{s,i} +A_{s,i} \sigma_{b,i}^2 (1+A_{s,i}/A_{b,i})\right]^{1/2},
\end{equation}
where $A_{s,i}$ and $A_{b,i}$ are the areas of the $i$th source and background apertures in pixels.
The scatter of the 25 (F110W) and 20 (F160W) net count measurements is found to be mainly due to the background nonuniformity. The uncertainty due to the source centroid is negligible.
In Table~\ref{tab:fluxesA}, we list the averages of these (aperture-corrected) fluxes, $N_S=\overline{{N}_{s,i}}$. The listed flux error encompasses the maximum flux uncertainty of the individual measurements ($\sigma_{N_{s,i}}$) and the variance of the 25 (F110W) and 20 (F160W) flux measurements, $\sigma_{N_{S}}$, as additional systematic error.\\ 
\begin{deluxetable}{ccccccccccccc}[h!]
\tablecolumns{15}
\tablecaption{Pulsar photometry with a few non-overlapping background apertures\label{tab:fluxesA}}
\tablewidth{0pt}
\tablehead{
\colhead{Band} & \colhead{$\lambda_{\rm piv}$} & \colhead{$\phi$} & \colhead{${\cal P}_\nu$} & \colhead{$t_{\rm exp}$} & 
\colhead{${\cal N}_s$} & \colhead{${\cal N}_b$} & \colhead{$\overline{N_{t,i}}$} & 
\colhead{$\overline{\sigma_{b,i}}$} & 
\colhead{${N}_{S}$}  & \colhead{$\sigma_{N_S}$} & 
\colhead{${f_{\nu}}$} & \colhead{STmag} \\  
\colhead{} & \colhead{$\mu$m} & \colhead{} & \colhead{nJy s/cnt} & \colhead{sec} &
\colhead{} & \colhead{} & \colhead{cnt} & \colhead{cnt/pix} & \colhead{cnt} & \colhead{cnt} &
\colhead{nJy}  & \colhead{ } 
}
\startdata
F110W & 1.153 & 0.739 &  68  &  648 & 5 & 5 & $3161$ & 2.9 & $1457$ & 289 & $207\pm 42$ & $27.22 \pm 0.22$\\
F160W & 1.537 & 0.608 & 153  & 1998 & 5 & 4 & $4621$ & 3.3 & $2089$ &  83 & $263\pm 16$ & $27.59 \pm 0.06$\\
\enddata
\tablecomments{
$\phi$ is the encircled count fraction in the 0\farcs2 radius source aperture,
${\cal P}_\nu$ is the filter's inverse sensitivity,
${\cal N}_s$ and ${\cal N}_b$
are the numbers of considered source and background apertures, all with an area of 122 pixels;
$\overline{N_{t,i}}$ and $\overline{N_{s,i}}$ are the averaged total (source and background) and net (source) number of counts,
$\sigma_{N_S}$ is
the variance of the background-subtracted source counts ($N_{s,i}$) for the 25 (F110W) and 20 (F160W) measurements, 
$\overline{\sigma_{b,i}}$ 
is the average of the rms in the background apertures, 
${f_{\nu}}$ 
is the averaged flux density at the pivot wavelength $\lambda_{\rm piv}$, where indiviudal fluxes are calculated as 
$f_{\nu,i}  = {\cal P}_\nu N_{s,i} (\phi \, t_{\rm exp})^{-1}$. 
STmag $= -21.10 -2.5\log \overline{f_\lambda}$, where the averaged flux density ${f_\lambda,i}$ is in units of erg cm$^{-2}$ s$^{-1}$ \AA$^{-1}$.}
\end{deluxetable}

For our second method, we use one source aperture, but increase the number of background apertures to improve the quality of the respective noise contribution estimate. Because of the inhomogeneous background, the choice of suitable aperture area is limited, and we 
use 30 (22) overlapping circular background apertures 
placed around the source in  F110W (F160W), as shown in the two right panels of Figure \ref{fig:apertures}. We utilize the same area for all source and background apertures ($0\farcs2$ radius).
Following the ``empty aperture'' approach \citep{Abramkin2022,Skelton2014}, we use multiple background aperture measurements to derive the average number of background counts, $\overline{N}_b$, and the variance $\sigma^2_{N_b}$. 
The total number of counts in the source aperture, $N_t$, is used to determine the net counts $N_s=N_t - \overline{N}_b$. 
The uncertainty of the net counts, $\sigma_{N_s}$, is then estimated to consist of contributions from the net counts themselves (estimated as a Poisson error) and the background noise contribution as:
\begin{equation} 
\sigma_{N_s} = {\large( \sigma^2_{N_b} + N_s \large)}^{1/2}.
\end{equation}
Table~\ref{tab:fluxesB} lists the values of these quantities as well as the aperture-corrected fluxes that are derived with the same exposure times and instrument-specific constants as in Table~\ref{tab:fluxesA}.
The aperture-corrected fluxes in each band from the two methods agree within $1\sigma$ for both bands. For further analysis, we choose the values from the ``empty aperture'' method.  

\begin{deluxetable}{ccccccccccc}[h!]
\tablecolumns{15}
\tablecaption{Pulsar photometry with many background apertures\label{tab:fluxesB}}
\tablewidth{0pt}
\tablehead{
\colhead{Band} &  \colhead{$\lambda_{\rm piv}$} & \colhead{$\phi$} & \colhead{${\cal P}_\nu$} & \colhead{$t_{\rm exp}$} &
\colhead{${N_t}$} & \colhead{${\cal N}_b$} & \colhead{$\overline{N}_b \pm \sigma_{N_b}$} & 
\colhead{$N_s$} & \colhead{${f_{\nu}}$} & \colhead{STmag} \\  
\colhead{} & \colhead{$\mu$m} & \colhead{} & \colhead{nJy s/cnt} & \colhead{sec} &  \colhead{cnt} & \colhead{} & \colhead{cnt} & \colhead{cnt} & 
\colhead{nJy}  & \colhead{ } 
}
\startdata
F110W & 1.153 & 0.739 &  68  &  648 & $3173 $ & 30 & $1567 \pm 282$ & $1606 \pm 285$ & $228\pm 40$ & $27.12 \pm 0.19$\\
F160W & 1.537 & 0.608 & 153  & 1998 & $4624 $ & 22 & $2553 \pm 125$ & $2071 \pm 133$ & $261\pm 17$ & $27.60 \pm 0.07$\\
\enddata
\tablecomments{
${N}_t$ indicates the total (source and background) counts in the source aperture.
${\cal N}_b$ background apertures are considered.
$\overline{N}_b$ and $\sigma_{N_b}$ are
the mean and variance of these ${\cal N}_b$ background measurements.
${N}_s = (N_t -\overline{N}_b) \pm \sigma_{N_s}$ lists the net (source) number of counts, with uncertainty $\sigma_{N_s}=(\sigma^2_{N_b}+ N_s)^{1/2}$.
The area of each aperture is 122 pixels.
Exposure times ($t_{\rm exp}$), the filter's inverse sensitivities (${\cal P}_\nu$) at the respective pivot wavelengths ($\lambda_{\rm piv}$), and the aperture corrections ($\phi$) are used in the same way as in Table~\ref{tab:fluxesA} to calculate aperture-corrected flux densities (${f_{\nu}}$) and the corresponding STmag.}
\end{deluxetable}

\end{document}